\documentclass[aps,prd,preprintnumbers,groupedaddress,nofootinbib,amssymb,eqsecnum,notitlepage]{revtex4-1}
\usepackage{graphicx}
\usepackage{bm}
\usepackage{amsmath}
\usepackage{color}
\usepackage{amsfonts}

\usepackage{graphicx}
\usepackage{bm}
\usepackage{amsmath}
\usepackage{color}
\usepackage{amsfonts}

\newcommand{\newc}{\newcommand}

\newcommand{\ben}{\begin{eqnarray}}
\newcommand{\een}{\end{eqnarray}}

\newc{\be}{\begin{equation}}
\newc{\ee}{\end{equation}}
\newc{\ba}{\begin{eqnarray}}
\newc{\ea}{\end{eqnarray}}
\newc{\bea}{\begin{eqnarray*}}
\newc{\eea}{\end{eqnarray*}}
\newc{\ie}{{\it i.e.} }
\newc{\eg}{{\it e.g.} }
\newc{\etc}{{\it etc.} }
\newc{\etal}{{\it et al.}}

\newc{\ra}{\rightarrow}
\newc{\lra}{\leftrightarrow}
\newc{\lsim}{\buildrel{<}\over{\sim}}
\newc{\gsim}{\buildrel{>}\over{\sim}}
\newc{\C}{{\cal C}}
\newc{\D}{{\cal D}}
\newc{\Xp}{X_{\phi}}
\newc{\Xc}{X_{\chi}}
\newc{\Mpl}{M_{\rm pl}}

\begin{document}
\baselineskip=12pt

\preprint{YITP-16-55} 
\preprint{IPMU16-0070}

\title{Effective gravitational couplings for cosmological perturbations \\
in generalized Proca theories}

\author{Antonio De Felice$^{1}$,
Lavinia Heisenberg$^{2}$,
Ryotaro Kase$^{3}$,
Shinji Mukohyama$^{1,4}$,\\
Shinji Tsujikawa$^{3}$ and
Ying-li Zhang$^{5,6}$}

\affiliation{$^1$Center for Gravitational Physics, Yukawa Institute for Theoretical Physics, Kyoto University, 606-8502, Kyoto, Japan\\
$^2$Institute for Theoretical Studies, ETH Zurich, Clausiusstrasse 47, 8092 Zurich, Switzerland\\
$^3$Department of Physics, Faculty of Science, Tokyo University of Science, 1-3, Kagurazaka,
Shinjuku-ku, Tokyo 162-8601, Japan\\
$^4$Kavli Institute for the Physics and Mathematics of the Universe (WPI),
The University of Tokyo Institutes for Advanced Study, The University of Tokyo,
Kashiwa, Chiba 277-8583, Japan\\
$^5$National Astronomy Observatories, Chinese Academy of Science,
Beijing 100012, People's Republic of China\\
$^6$Institute of Cosmology and Gravitation,
University of Portsmouth, Portsmouth PO1 3FX, UK}

\date{\today}

\begin{abstract}
We consider the finite interactions of the generalized Proca theory 
including the sixth-order Lagrangian and derive the full linear perturbation 
equations of motion on the flat Friedmann-Lema\^{i}tre-Robertson-Walker background
in the presence of a matter perfect fluid.
By construction, the propagating degrees of freedom (besides the matter perfect fluid) 
are two transverse vector perturbations, one longitudinal scalar,
and two tensor polarizations. 
The Lagrangians associated with intrinsic
vector modes neither affect the background equations of motion
nor the second-order action of tensor perturbations, but they do give
rise to non-trivial modifications to the no-ghost condition of vector perturbations
and to the propagation speeds of vector and scalar perturbations.
We derive the effective gravitational coupling $G_{\rm eff}$ with
matter density perturbations under a quasi-static approximation
on scales deep inside the sound horizon.
We find that the existence of intrinsic vector modes allows
a possibility for reducing $G_{\rm eff}$. In fact, within the parameter space, $G_{\rm eff}$
can be even smaller than the Newton gravitational constant $G$
at the late cosmological epoch, with a peculiar phantom dark energy
equation of state (without ghosts). The modifications to the slip parameter $\eta$
and the evolution of growth rate $f\sigma_8$ are discussed as well.
Thus, dark energy models in the framework of generalized Proca theories
can be observationally distinguished from the $\Lambda$CDM model
according to both cosmic growth and expansion history. Furthermore,
we study the evolution of vector perturbations and show that outside
the vector sound horizon the perturbations are nearly frozen and start to
decay with oscillations after the horizon entry.

\end{abstract}


\maketitle

\section{Introduction}

The discovery of a late-time acceleration of the Universe \cite{SNIa}
pushed forward an idea that one or more additional degrees of freedom (DOF)
to those appearing in the standard model of particle physics may be
the origin of dark energy \cite{review}.
The simplest example is a minimally coupled scalar field dubbed
``quintessence'' \cite{quin}. The cosmic acceleration can be realized
for the scalar field with a slowly varying potential, in which case
the dark energy equation of state $w_{\rm DE}$ is larger than $-1$.
The cosmological constant can be regarded as the non-propagating
limit of quintessence (i.e., vanishing kinetic energy) with $w_{\rm DE}=-1$.
The likelihood analysis based on Supernovae type Ia (SN Ia),
Cosmic Microwave Background (CMB), and Baryon Acoustic
Oscillations (BAO) showed no statistically significant signatures that
quintessence is observationally favored over
the cosmological constant \cite{Chiba}.

There are models of dark energy in which the scalar field $\phi$ has
a non-minimal coupling to the Ricci scalar $R$ with the form
$F(\phi)R$, where $F(\phi)$ is a function of $\phi$ \cite{stensor}.
Brans-Dicke theory \cite{Brans} with a scalar potential is one of the
examples for such modified gravitational theories.
For dark energy models in the framework of non-minimally coupled
theories it is possible to realize $w_{\rm DE}$ smaller
than $-1$ \cite{stensorw,fRw} without ghosts.
Since the gravitational interaction is also different from that in
General Relativity (GR), these models leave several interesting
observational signatures that can be distinguished from
the $\Lambda$-Cold-Dark-Matter ($\Lambda$CDM) model \cite{obsersig}.

The non-minimal coupling $F(\phi)R$
can be extended to contain a derivative coupling in the form
$F(\phi,X)R$, where $X=-\partial_{\mu}\phi \partial^{\mu}\phi/2$
is the field kinetic energy. In general, unless some counter terms
are introduced, such derivative couplings give rise to the equations
of motion higher than second order \cite{Deffayet09}.
The appearance of time derivatives higher than two leads to the
so-called Ostrogradski instability \cite{Ostro} with the Hamiltonian
unbounded from below.
In 1974 Horndeski derived most general scalar-tensor theories
with second-order equations of motion \cite{Horndeski},
which received much attention over the past five years
in connection to the problems of dark energy and inflation \cite{Horn2}.
A sub-class of Horndeski interactions also naturally arise in massive gravity \cite{deRham:2011by}.
In scalar-tensor Horndeski theories, there is one scalar propagating
DOF besides the two tensor polarizations.

If we consider a vector field $A^{\mu}$ as the source of dark energy,
the number of DOF generally increases relative to scalar-tensor
Horndeski theories. The massless Maxwell field given by the Lagrangian
${\cal L}_F=-F^{\mu \nu}F_{\mu \nu}/4$ (where
$F^{\mu \nu}=\partial^{\mu}A^{\nu}-\partial^{\nu}A^{\mu}$)
has two transverse polarizations of the vector mode
with a protected $U(1)$ gauge symmetry.
Introduction of the vector mass term gives rise to
the additional longitudinal propagation of
a scalar mode due to the breaking of gauge invariance.
In GR with the massive Proca field,
there are two transverse and one longitudinal propagating
DOF besides the two tensor polarizations.

For the massless gauge-invariant vector field coupled to gravity
with Lorentz symmetry, there is a no-go theorem stating
that derivative interactions similar to those appearing for
covariant Galileons \cite{coGa} do not arise for a single spin-$1$ field in any
dimensions \cite{Gum,Deffayet:2016von}
(see also Ref.~\cite{Deffayet:2010zh}).
This situation is different for massive Proca theories in which
the $U(1)$ gauge invariance is explicitly broken.
Analogous to scalar-tensor Horndeski theories,
it is possible to construct an action of generalized Proca
theories with second-order equations of motion having three
propagating DOF with two tensor
polarizations. The corresponding action has been constructed by using
the Levi-Civita tensor to avoid the appearance of time
derivatives higher than two. In fact, the analysis based on the
Hessian matrix showed that such theories do not propagate extra DOF
other than those mentioned above \cite{Heisenberg}.
A sub-class of these interactions was also discussed in \cite{Tasinato}.

If we impose the condition that the scalar part of the vector field
only has terms that do not correspond to trivial total derivative interactions,
then the series of the generalized Proca Lagrangian
stops at quintic order (${\cal L}_5$) \cite{Heisenberg}. By relaxing this condition,
it is also possible to construct higher-order derivative interactions
associated with the intrinsic vector part \cite{Allys,Jimenez16}.
The sixth-order Lagrangian ${\cal L}_6$ \cite{Jimenez16},
which contains the double dual Riemann tensor, accommodates 
an interaction term in the gauge-invariant 
vector-tensor theories constructed by Horndeski in 1976 \cite{Horndeskivec}.
In Ref.~\cite{Allys} the authors derived seventh and higher-order derivative
interactions having two transverse and one longitudinal polarizations,
but it was later found that they correspond to trivial interactions
by virtue of the Cayley-Hamilton theorem.
Thus, it suffices to consider the Lagrangians up to sixth order
presented in Ref.~\cite{Jimenez16}.

Recently, the cosmology in generalized Proca
theories up to the quintic Lagrangian ${\cal L}_5$ was
studied in Ref.~\cite{DeFelice16} (see also
Refs.~\cite{Barrow,Jimenez,TKK,Hull,Li,Jimenez:2014rna}
for earlier related works).
In such theories, there is a non-trivial branch of
the background solutions where
the temporal vector component $\phi$ depends on the
Hubble expansion rate $H$ alone. In Ref.~\cite{DeFelice16}
the authors proposed a dark energy model in which the solutions
finally approach a de Sitter attractor characterized by constant $\phi$.
The conditions for avoiding ghosts and Laplacian
instabilities were generally derived for tensor, vector, and scalar
perturbations, which were applied to the proposed dark energy 
model to search for theoretically consistent parameter spaces. 
Moreover, there exists viable model parameter spaces in which 
the propagation speed of tensor perturbations is consistent with the 
Cherenkov-radiation constraint \cite{Cherenkov} and the recent detection 
of gravitational waves \cite{GWde}.
In addition, the cubic and quartic derivative interactions 
allow the screening of the fifth force mediated by the vector
field \cite{scvector}.

In this paper, we extend the analysis of Ref.~\cite{DeFelice16} to include
the sixth-order Lagrangian ${\cal L}_6$ as well as the quadratic Lagrangian
${\cal L}_2$ containing the dependence of $X=-A_{\mu} A^{\mu}/2$,
$F=-F_{\mu \nu} F^{\mu \nu}/4$, and
$Y=A^{\mu}A^{\nu} {F_{\mu}}^{\alpha}F_{\nu \alpha}$
(which accommodates the terms discussed in
Ref.~\cite{Fleury}).
We derive full linear perturbation equations of motion for tensor, vector, and
scalar modes at linear order in the presence of a perfect fluid
and then obtain the effective gravitational coupling
$G_{\rm eff}$ with matter by employing
a quasi-static approximation for perturbations deep inside
the sound horizon.
We also study the growth rate of matter perturbations and
the evolution of gravitational potentials to confront generalized
Proca theories with the observations of
redshift-space distortions (RSD), CMB, and weak lensing.

The recent observations of RSD \cite{Beu,Ledo,Eriksen} and
cluster counts \cite{Vik} have shown
that the cosmic growth rate is lower than that
predicted by the $\Lambda$CDM model with
$\sigma_8$ constrained by the Planck CMB data \cite{Planck}.
This tension reduces with the WMAP bound on $\sigma_8$ \cite{WMAP}
and the systematic errors of RSD data are still quite large.
Hence, in current observations, one cannot conclusively say that
weak gravity is really favored over the gravitational law of GR.
However, it is of interest to look for the theoretical possibility of realizing
weak gravity on cosmological scales.
In scalar-tensor Horndeski theories, unless the second-order action of tensor perturbations
is modified from GR to a large extent, it is difficult to realize $G_{\rm eff}<G$
without ghosts due to the presence of attractive scalar-matter
couplings \cite{Tsuji15} 
(see also 
Refs.~\cite{DKT,Amendola,Bellini,Perenon,deRham:2014naa,deRham:2014fha}).
It remains to see whether the existence of the vector field
can modify this situation. We shall pursue the possibility of weak gravity
for a class of dark energy models in generalized Proca theories.

This paper is organized as follows.
In Sec.~\ref{HPsec} we obtain the background equations of motion in the presence of
a perfect fluid containing the generalized Proca Lagrangian up to sixth order.
In Sec.~\ref{spsec} we derive the equations of motion
for tensor and vector perturbations and identify no-ghost and
stability conditions of them in the small-scale limit.
In Sec.~\ref{scasec} the scalar perturbation equations
and the observables associated with large-scale structures,
CMB, and weak lensing will be discussed.
In Sec.~\ref{quasisec} we analytically obtain the effective gravitational
coupling with matter perturbations under the quasi-static approximation
and derive a necessary condition for realizing $G_{\rm eff}<G$.
In Sec.~\ref{obsersec} we study the evolution of observable
quantities for dark energy models in a class of generalized Proca
theories and discuss how the vector field affects $G_{\rm eff}$.
Sec.~\ref{consec} is devoted to conclusions.

\section{Generalized Proca theories and the background
equations of motion}
\label{HPsec}

We study generalized Proca theories with two transverse and
one longitudinal polarizations of a vector field $A^{\mu}$
coupled to gravity.
The action of such theories is of the
following forms \cite{Heisenberg,Jimenez16}
\be
S=\int d^4x \sqrt{-g} \left( {\cal L}
+{\cal L}_M \right)\,,\qquad
{\cal L}=\sum_{i=2}^{6} {\cal L}_i\,,
\label{Lag}
\ee
where $g$ is a determinant of the metric tensor
$g_{\mu \nu}$, ${\cal L}_M$ is a matter
Lagrangian, and ${\cal L}_{2,3,4,5,6}$
are given by
\ba
{\cal L}_2 &=& G_2(X,F,Y)\,,
\label{L2}\\
{\cal L}_3 &=& G_3(X) \nabla_{\mu}A^{\mu}\,,
\label{L3}\\
{\cal L}_4 &=&
G_4(X)R+
G_{4,X}(X) \left[ (\nabla_{\mu} A^{\mu})^2
-\nabla_{\rho}A_{\sigma}
\nabla^{\sigma}A^{\rho} \right]\,,\label{L4} \\
{\cal L}_5 &=&
G_{5}(X) G_{\mu \nu} \nabla^{\mu} A^{\nu}
-\frac16 G_{5,X}(X) [ (\nabla_{\mu} A^{\mu})^3
-3\nabla_{\mu} A^{\mu}
\nabla_{\rho}A_{\sigma} \nabla^{\sigma}A^{\rho}
+2\nabla_{\rho}A_{\sigma} \nabla^{\gamma}
A^{\rho} \nabla^{\sigma}A_{\gamma}] \nonumber \\
& &-g_5(X) \tilde{F}^{\alpha \mu}
{\tilde{F^{\beta}}}_{\mu} \nabla_{\alpha} A_{\beta}\,,
\label{L5}\\
{\cal L}_6 &=& G_6(X) L^{\mu \nu \alpha \beta}
\nabla_{\mu}A_{\nu} \nabla_{\alpha}A_{\beta}
+\frac12 G_{6,X}(X) \tilde{F}^{\alpha \beta} \tilde{F}^{\mu \nu}
\nabla_{\alpha}A_{\mu} \nabla_{\beta}A_{\nu}\,,
\label{L6}
\ea
with $F_{\mu \nu}=\nabla_{\mu}A_{\nu}-\nabla_{\nu}A_{\mu}$
(and $\nabla_{\mu}$ is the covariant
derivative operator).
The function $G_2$ depends on the following
three quantities
\ba
X &=&-\frac12 A_{\mu} A^{\mu}\,,\\
F &=& -\frac14 F_{\mu \nu} F^{\mu \nu}\,,\\
Y &=& A^{\mu}A^{\nu} {F_{\mu}}^{\alpha}
F_{\nu \alpha}\,,
\label{Xdef}
\ea
whereas $G_{3,4,5,6}$ and $g_5$ are arbitrary
functions of $X$ with the notation of partial
derivatives as $G_{i,X} \equiv \partial G_{i}/\partial X$.
The vector field is coupled to the Ricci scalar $R$ and
the Einstein tensor $G_{\mu \nu}$
through the functions $G_4(X)$ and $G_5(X)$\footnote{It would be interesting to study the consequences of the vector field living on a composite effective metric as it could be for instance the case in massive gravity \cite{deRham:2014naa}. This will be studied in a future work.}.
The $L^{\mu \nu \alpha \beta}$ and
$\tilde{F}^{\mu \nu}$ are the double dual Riemann
tensor and the dual strength tensor
defined, respectively, by
\be
L^{\mu \nu \alpha \beta}=\frac14 \epsilon^{\mu \nu \rho \sigma}
\epsilon^{\alpha \beta \gamma \delta} R_{\rho \sigma \gamma \delta}\,,
\qquad
\tilde{F}^{\mu \nu}=\frac12 \epsilon^{\mu \nu \alpha \beta}
F_{\alpha \beta}\,,
\ee
where $\epsilon^{\mu \nu \rho \sigma}$ is the
Levi-Civita tensor and $R_{\rho \delta \gamma \delta}$
is the Riemann tensor.

In the original Proca theory on the Minkowski background,
which corresponds to the functions $G_2(X)=m^2 X$ and
$G_{3,4,5,6}=0$, the $U(1)$ gauge symmetry is explicitly
broken due to the non-vanishing mass $m$ of the vector field.
In this case, the longitudinal mode arises in addition
to the two transverse polarizations.
The Lagrangians given above are the generalization of
Proca theories coupled to gravity in which the number of propagating DOF remains three besides the two graviton polarizations.
The existence of non-minimal couplings in ${\cal L}_{4,5,6}$
comes from the demand for keeping the three propagating DOF
with second-order equations of motion.
The gauge-invariant vector-tensor interaction introduced
by Horndeski in 1976 corresponds to the Lagrangian
${\cal L}=F+{\cal L}_4+{\cal L}_6$ with
constant functions $G_4$ and $G_6$ \cite{Horndeskivec}.

In Ref.~\cite{Heisenberg} there exists a term of the form
$f_4(X)(\nabla_{\rho}A_{\sigma}
\nabla^{\rho}A^{\sigma}-\nabla_{\rho}A_{\sigma}
\nabla^{\sigma}A^{\rho})$ with $f_4(X)=c_2G_{4,X}$
in the Lagrangian ${\cal L}_4$, but it can be expressed in terms of $X$ and $F$ as $-2f_4(X)F$. Hence such a term has been absorbed into the Lagrangian ${\cal L}_2$.
The term multiplied by $d_2G_{5,X}(X)$ in the Lagrangian
${\cal L}_5$ of Ref.~\cite{Heisenberg}, which corresponds
to an intrinsic vector mode, is now replaced with
the last contribution in Eq.~(\ref{L5}).
The function $g_5(X)$ does not need to have a relation
with $G_{5,X}(X)$ \cite{Allys,Jimenez16}, so the prescription in this paper
is more general than that of Ref.~\cite{Heisenberg}. Furthermore,
we adapt to the same notation as in Ref.~\cite{Jimenez16}, 
which agrees completely with Ref.~\cite{Allys}.

In the Lagrangian ${\cal L}_2$, we have also taken into
account the dependence of the quantity $Y$ that can be
constructed from $A^{\mu}$ and its derivatives
up to first order \cite{Heisenberg,Fleury}.
In principle we can also include the dependence of
the term $F^{\mu \nu} \tilde{F}_{\mu \nu}$
in ${\cal L}_2$. If we impose the parity invariance,
however, such a term is irrelevant to the perturbations
at linear order. Hence we shall consider the function $G_2$
depending on the three quantities $X, F, Y$ in this paper.

Let us consider the flat
Friedmann-Lema\^{i}tre-Robertson-Walker
(FLRW) background described with the line element
$ds^2=-dt^2+a^2(t)d{\bm x}^2$, where $a(t)$
is the time-dependent scale factor.
To keep the spatial isotropy of the background, the
vector field needs to have a time-dependent
temporal component $\phi(t)$ alone, i.e.,
\be
A^{\mu}=(\phi(t),0,0,0)\,.
\ee
For the matter Lagrangian ${\cal L}_M$ we consider
a perfect fluid with the energy density $\rho_M$ and
the isotropic pressure $P_M$.
Assuming that matter is minimally coupled to gravity,
we have the continuity equation
\be
\dot{\rho}_M+3H(\rho_M+P_M)=0\,,
\label{continuity}
\ee
where a dot denotes a derivative with respect to $t$, and
$H \equiv \dot{a}/a$ is the expansion rate of the Universe.

Variation of the action (\ref{Lag}) with respect to
$g_{\mu \nu}$ leads to the background equations of motion
\ba
& &
G_2-G_{2,X}\phi^2-3G_{3,X}H \phi^3
+6G_4H^2-6(2G_{4,X}+G_{4,XX}\phi^2)H^2\phi^2
+G_{5,XX} H^3\phi^5+ 5G_{5,X} H^3\phi^3
=\rho_M\,,
\label{be1}\\
& &
G_2-\dot{\phi}\phi^2G_{3,X}+2G_4\,(3H^2+2\dot{H})
-2G_{4,X}\phi \, ( 3H^2\phi +2H\dot{\phi}
+2\dot{H} \phi )-4G_{4,XX}H\dot{\phi}\phi^3\nonumber\\
&& {}+G_{5,XX}H^2\dot{\phi} \phi^4+G_{5,X}
H \phi^2(2\dot{H}\phi +2H^2\phi+3H\dot{\phi})
=-P_M\,.
\label{be2}
\ea
Varying the action (\ref{Lag}) with respect to $A^{\mu}$,
it follows that
\be
\phi \left( G_{2,X}+3G_{3,X}H\phi +6G_{4,X}H^2
+6G_{4,XX}H^2\phi^2
-3G_{5,X}H^3\phi-G_{5,XX}H^3 \phi^3 \right)=0\,.
\label{be3}
\ee
Equations (\ref{be1})-(\ref{be3}) are exactly
the same as those derived for more specific theories
containing the Lagrangians up to
${\cal L}_5$ \cite{DeFelice16}.
Hence the Lagrangian ${\cal L}_6$ and the dependence of
$F$ and $Y$ in ${\cal L}_2$ do not affect the background equations.
In Eq.~(\ref{be3}) there exists a branch with $\phi \neq 0$,
which gives rise to interesting de Sitter solutions
characterized by constant $\phi$ and $H$ \cite{DeFelice16}.

\section{Tensor and vector perturbations}
\label{spsec}

In what follows we derive the equations of motion
for tensor, vector, and scalar perturbations on the
flat FLRW background.
The discussions about scalar perturbations
will be given separately in Sec.~\ref{scasec}.

First of all, we decompose temporal and spatial components
of the vector field $A^{\mu}(t,{\bm x})$ into the
background and perturbed components, as
\ba
A^{0} & = & \phi(t)+\delta\phi\,,\\
A^{i} & = & \frac{1}{a^2(t)} \delta^{ij} \left(
\partial_{j}\chi_{V}+E_j \right)\,,
\ea
where the perturbation $\delta \phi$ depends on
$t$ and ${\bm x}$.
The perturbations $\chi_V$ and $E_j$
correspond to the intrinsic scalar and vector parts,
respectively, where the latter satisfies the transverse
condition $\partial^jE_j=0$.

As for the matter sector, we consider a single perfect fluid described
by the Schutz-Sorkin action \cite{Sorkin}:
\be
S_{M}=-\int d^{4}x \left[ \sqrt{-g}\,\rho_M(n)
+J^{\mu}(\partial_{\mu}\ell+\mathcal{A}_1
\partial_{\mu}\mathcal{B}_1+\mathcal{A}_2
\partial_{\mu}\mathcal{B}_2) \right],
\label{Spf}
\ee
where the fluid energy density $\rho_M$ depends on its number density defined by
\be
n=\sqrt{\frac{J^{\alpha}J^{\beta}g_{\alpha\beta}}{g}}\,,
\label{num}
\ee
and $J^\mu$ is a vector field of weight one,
$\ell$ is a scalar, $\mathcal{A}_{1,2}, \mathcal{B}_{1,2}$
are scalar quantities associated with vector perturbations.

On the FLRW background the temporal component $J^0$ corresponds to the total fluid number ${\cal N}_0$,
which is constant.
{}From Eq.~(\ref{num}) the background number
density $n_0$ reads
\be
n_0=\frac{{\cal N}_0}{a^3}\,.
\ee
The temporal component $\partial_0 \ell$ is equivalent to
$-\rho_{M,n} \equiv -\partial\rho_M/\partial n$ at the background
level, so that the matter action (\ref{Spf}) reduces to
\be
S_M^{(0)}=\int d^4 x \sqrt{-g}\, P_M\,, \qquad
P_M=n_0\rho_{M,n}-\rho_M\,,
\label{PM}
\ee
where $P_M$ corresponds to the pressure of the perfect fluid.

The scalar quantities $J^{0}$ and $\ell$ have the perturbations
$\delta J$ and $v$, respectively, so they can be written as
\ba
J^{0} &=&  \mathcal{N}_{0}+\delta J\,,\\
\ell &=&-\int^{t} \rho_{M,n} dt'
-\rho_{M,n}\,v\,,
\label{elldef}
\ea
where $v$ corresponds to the velocity potential.
One can express the spatial components of $J^{\mu}$
in terms of the sum of the scalar part $\delta j$ and
the vector part $W_k$, as
\be
J^{i} =\frac{1}{a^2}\,\delta^{ik}
\left( \partial_{k}\delta j+W_k \right)\,.
\ee
The vector perturbation $W_k$ obeys the transverse
condition $\partial^k W_k=0$.
If we consider the vector field in the form
$W_k=(W_1(t,z),W_2(t,z),0)$ whose
$x$ and $y$ components depend on $t$ and $z$ alone,
then it automatically satisfies the transverse condition.
For the quantities ${\cal A}_i$ and ${\cal B}_i$ appearing
in Eq.~(\ref{Spf}), the simplest choice keeping the
required property of the vector mode is given by \cite{DGS}
\ba
& &
{\cal A}_1=\delta {\cal A}_1(t,z)\,,\qquad  \quad~
{\cal A}_2=\delta {\cal A}_2(t,z)\,,\nonumber \\
& &
{\cal B}_1=x+\delta {\cal B}_1(t,z)\,,\qquad
{\cal B}_2=y+\delta {\cal B}_2(t,z)\,,\label{ABi}
\ea
where $\delta {\cal A}_i$ and $\delta {\cal B}_i$ are perturbed quantities.

On taking variation of the matter action with respect to the field $J^\mu$, 
we find the fluid normalized four-velocity $u_\mu$ as
\be
u_\mu\equiv\frac{J_\mu}{n\sqrt{-g}}=\frac1{\rho_{M,n}}\,(\partial_{\mu}\ell+\mathcal{A}_1
\partial_{\mu}\mathcal{B}_1+\mathcal{A}_2\partial_{\mu}\mathcal{B}_2)\,,
\label{umudef}
\ee
whose spatial components, on the FLRW manifold, are split, in terms of 
scalar and vector perturbations, as
\be
u_i=-\partial_i v+v_i\,,
\label{eq:uVectPert}
\ee
where $v$ is the velocity potential given in Eq.~(\ref{elldef}) and 
$v_i$ is a transverse three-vector satisfying $\partial^i v_i=0$.
{}From Eqs.~(\ref{ABi}) and (\ref{umudef}) the intrinsic vector part $v_i$ 
is related with the linear perturbation $\delta {\cal A}_i$, as
$\delta {\cal A}_i=\rho_{M,n}v_i$. 
The equation of motion for $\delta {\cal A}_i$ follows
by varying the second-order action of the vector
field with respect to the perturbation $\delta {\cal B}_i$.

For the gravity sector, we consider the linearly perturbed
line-element in the flat
gauge \cite{Bardeen,KS,Mukhanov,BTW}:
\be
ds^{2}=-(1+2\alpha)\,dt^{2}+2\left( \partial_{i}\chi
+V_i \right)dt\,dx^{i}+a^{2}(t) \left( \delta_{ij}
+h_{ij} \right)dx^i dx^j\,,
\ee
where $\alpha, \chi$ are scalar metric perturbations,
$V_i$ is the vector perturbation obeying the transverse
condition $\partial^i V_i=0$, and $h_{ij}$ is
the tensor perturbation satisfying the transverse and
traceless conditions $\partial^i h_{ij}=0$ and
${h_i}^i=0$. The temporal and spatial components
of gauge transformation vectors are completely
fixed under the above gauge choice.

\subsection{Tensor perturbations}

We can express the tensor perturbation $h_{ij}$ in terms of
the two polarization modes $h_{+}$ and $h_{\times}$, as
$h_{ij}=h_{+}e_{ij}^{+}+h_{\times} e_{ij}^{\times}$,
where $e_{ij}^{+}$ and $e_{ij}^{\times}$ obey the relations
$e_{ij}^{+}({\bm k}) e_{ij}^{+}(-{\bm k})^*=1$,
$e_{ij}^{\times}({\bm k}) e_{ij}^{\times}(-{\bm k})^*=1$,
and $e_{ij}^{+}({\bm k}) e_{ij}^{\times}(-{\bm k})^*=0$
in Fourier space (${\bm k}$ is the comoving wave number).
The second-order action for tensor perturbations, which
is derived after expanding Eq.~(\ref{Lag}) in $h_{ij}$ up
to quadratic order, reads
\be
S_T=\sum_{\lambda={+},{\times}}\int dt\,d^3x\,
a^3\,\frac{q_T}{8}  \left[\dot{h}_\lambda^2
-\frac{c_T^2}{a^2}(\partial h_\lambda)^2\right]\,,
\label{ST}
\ee
where
\be
q_T=2G_4-2\phi^{2}G_{{4,X}}+
H\phi^{3}G_{{5,X}}\,,
\label{qT}
\ee
and
\be
c_{T}^{2}=
\frac {2G_{4}+\phi^{2}\dot\phi\,G_{{5,X}}}{q_T}\,.
\label{cT}
\ee

The quantities $q_T$ and $c_T^2$ are the same as
those derived in Ref.~\cite{DeFelice16}, so
the Lagrangian ${\cal L}_6$ and the terms
$F$ and $Y$ in ${\cal L}_2$ do not affect the dynamics
of tensor perturbations.
Varying the action (\ref{ST}) with respect to $h_{\lambda}$,
the tensor perturbation equation of motion in Fourier space
is given by
\be
\ddot{h}_{\lambda}+\left( 3H+
\frac{\dot{q}_T}{q_T} \right) \dot{h}_{\lambda}
+c_T^2 \frac{k^2}{a^2}h_{\lambda}=0\,,
\ee
where $k=|{\bm k}|$.
The tensor ghost and small-scale Laplacian instabilities
are absent for $q_T>0$ and $c_T^2>0$, respectively.

\subsection{Vector perturbations}
\label{vecsec}

As we have already mentioned, the vector perturbations $E_i$, $W_i$,
$\delta {\cal A}_i$, and $V_i$ obey the transverse conditions, so the
components of these fields can be chosen as
$E_i=(E_1(t,z),E_2(t,z),0)$.  On using Eq.~(\ref{ABi}) and expanding
the matter action (\ref{Spf}) up to quadratic order in vector
perturbations, the second-order action reads \cite{DeFelice16}
\be
(S_{M}^{(2)})_V=\int dtd^3x  \sum_{i=1}^{2}
\left[ \frac{1}{2a^2{\cal N}_0} \left\{
\rho_{M,n} \left (W_i^2+{\cal N}_0^2\, V_i^2 \right)
+{\cal N}_0 \left(2\rho_{M,n}V_iW_i-a^3\rho_M V_i^2 \right)
\right\}-{\cal N}_0 \delta {\cal A}_i \dot{\delta \cal B}_i
-\frac{1}{a^2} W_i \delta {\cal A}_i \right]\,,
\label{Lv}
\ee
where the quantities $W_i, \delta {\cal A}_i, \delta {\cal B}_i$
appear only in the matter action (\ref{Lv}) but not in
the quadratic action originating from
$\int d^4 x\sqrt{-g}{\cal L}$.

Varying Eq.~(\ref{Lv}) with respect to $W_i$, it follows that
\be
W_i=\frac{{\cal N}_0 (\delta {\cal A}_i-\rho_{M,n}V_i)}{\rho_{M,n}}
={\cal N}_0\,(v_i-V_i)\,.
\label{Wiex}
\ee
On using this relation and varying the matter action
with respect to $\delta {\cal A}_i$, we obtain
\be
\delta {\cal A}_i=\rho_{M,n}v_i\,,\quad
{\rm where} \quad
v_i=V_i-a^2\dot{\delta {\cal B}}_i\,.\label{eq:dotBi}
\ee
Similarly, variation of the matter action with respect to
$\delta {\cal B}_i$ gives rise to the conservation equation
\be
\rho_{M,n} v_{i}
=\frac{(\rho_M+P_M)}{n_0}\,v_{i}=\delta\mathcal{A}_i=C_i\,,
\label{conser}
\ee
where $C_i$ are two constants in time (but may be dependent on $k$), 
which are related to the initial conditions for the intrinsic vector modes in
the fluid. Therefore, the dynamics of $v_i$ is completely
determined as
\be
v_i = \frac{\mathcal{N}_0\,C_i}{(\rho_M+P_M)a^3}\,.
\label{eq:uDyn}
\ee
Then, after integrating out the fields $W_i$ and 
$\delta A_i$, the resulting second-order matter action reduces to
\be
(S_{M}^{(2)})_V=\int dtd^3x \sum_{i=1}^{2}
\frac{a}{2} \left[ \left( \rho_M+P_M \right)
\left( V_i-a^2\dot{\delta {\cal B}}_i \right)^2
-\rho_M V_i^2 \right]\,.
\label{LVM2}
\ee

To expand the action (\ref{Lag}) up to second order, 
it is convenient to introduce the following combination
\be
Z_i= E_i+\phi(t)\,V_i\,,
\ee
so that $A_i=Z_i$ for vector perturbations.
We also introduce the rescaled fields
\be
\tilde{V}_i \equiv \frac{1}{a}V_i\,,\qquad
\tilde{Z}_i \equiv \frac{1}{a}Z_i\,.
\ee
Taking into account Eq.~(\ref{LVM2}), the full quadratic 
action for vector perturbation reads
\ba
S_V^{(2)} &=& \int dtd^3x \sum_{i=1}^{2}a^3 \biggl[
\frac{q_V}{2} \dot{\tilde{Z}}_i^2
-\frac{1}{2a^2}{\cal C}_1 (\partial \tilde{Z}_i)^2
-\frac{1}{2}{\cal C}_2\tilde{Z}_i^2
+\frac{\phi}{2a^2} \left( 2G_{4,X}-G_{5,X}H \phi \right)
\partial \tilde{V}_i \partial \tilde{Z}_i \nonumber \\
& &\qquad \qquad \quad \quad~~
+\frac{q_T}{4a^2} (\partial \tilde{V}_i)^2
+\frac12 (\rho_M+P_M)(\tilde{V}_i-a\dot{\delta {\cal B}}_i )^2 \biggr]\,,
\label{Lv4}
\ea
where
\ba
q_V
&=& G_{2,F}+2G_{2,Y}\phi^2-4g_5H \phi
+2G_6H^2+2G_{6,X} H^2 \phi^2\,,\\
{\cal C}_1
&=& q_V + 2[G_6 \dot{H}-G_{2,Y}\phi^2
-(H\phi-\dot{\phi})(H \phi G_{6,X}-g_5)]\,,\\
{\cal C}_2
&=& 2( 2\,G_{{4,X}} -H\phi\,G_{{5,X}}) \dot{H} + ( G_{{3,X}}
+4\,\phi\,HG_{{4,XX}}-G_{{5,X}}{H}^{2}-{\phi}^{2}G_{{5,XX}}{H}^{2} )\dot\phi 
+2q_V H^2+\frac{d}{dt}(q_V H)\,.
\ea
Since $\tilde{V}_i$ is not coupled with $\dot{\tilde{Z}}_i$, 
the kinetic term of the field $\tilde{Z}_i$ remains unchanged 
after the integration of $\tilde{V}_i$.
Hence we need to impose the condition $q_V>0$ to avoid that 
$\tilde{Z}_i$ becomes a ghost field. 
The auxiliary fields $\dot{\delta{\cal B}}_i$
acquire a kinetic term which is trivially positive for $q_T>0$.

It should be noted that the dynamics of the vector perturbations is
completely determined by the initial conditions of $\tilde{Z}_i$ and
$\dot{\tilde{Z}}_i$ and by the two constants $C_i$. In fact, in Fourier space,
on using Eqs.\ (\ref{eq:dotBi}) and (\ref{eq:uDyn}), the equations of
motion for $\tilde{V}_{i}$ and $\tilde{Z}_i$ following from Eq.~(\ref{Lv4}) are given,
respectively, by
\ba
& &
\frac{q_T}{2}
\frac{k^{2}}{a^{2}}\tilde{V}_{i}=-\frac{\mathcal{N}_0\,C_i}{a^4}-\frac\phi2\,
\left(2G_{4,X}- G_{5,X}H \phi \right)\,\frac{k^{2}}{a^{2}}\tilde{Z}_{i}\,,\label{vecre}\\
& &
\ddot{\tilde{Z}}_i+\left(3 H+\frac{\dot{q}_V}{q_V} \right)\dot{\tilde{Z}}_i
+\left[\frac{{\cal C}_1}{q_V}+\frac{\phi^2}{2q_Vq_T}\,
\left(2G_{4,X}- G_{5,X}H \phi \right)^2 \right]\frac{k^2}{a^2}\tilde{Z}_i
+\frac{{\cal C}_2}{q_V} \tilde{Z}_i\nonumber\\
&&=-\frac{\phi}{q_Vq_T} 
\left( 2G_{4,X}-G_{5,X}H \phi \right)\frac{\mathcal{N}_0\,C_i}{a^4}\,.
\label{Zieq}
\ea
This shows that there are only two dynamical fields $\tilde{Z}_1$ 
and $\tilde{Z}_2$
and that the matter fields can influence their dynamics only via the term
on the r.h.s.\ of Eq.\ (\ref{Zieq}), which is independent of the
matter equation of state. {}From Eq.~(\ref{Zieq}) we define the mass 
squared of the vector fields $\tilde{Z}_i$, as 
\be
m_V^2 \equiv \frac{\mathcal{C}_2}{q_V}\,.
\ee
We can easily see, from the expression of ${\cal C}_2$, that, 
on the de Sitter solution characterized by $\dot{H}=0$ 
and $\dot{\phi}=0$, $m_V^2$ reduces to $2H^2$.

The vector propagation speed squared $c_V^2$ corresponds to 
the coefficient in front of the $(k^2/a^2)\tilde{Z}_i$ term
in Eq.~(\ref{Zieq}), i.e., 
\be
c_{V}^{2}=1+\frac{\phi^2(2G_{4,X}-G_{5,X}H \phi)^2}
{2q_Tq_V}+\frac{2[G_6 \dot{H}-G_{2,Y}\phi^2
-(H\phi-\dot{\phi})(H \phi G_{6,X}-g_5)]}
{q_V}\,,
\label{cv}
\ee
which is required to be positive for the stability on 
small scales.
The Lagrangian ${\cal L}_6$,
the contribution $Y$ to ${\cal L}_2$, and the $g_5$-dependent term in
${\cal L}_5$ affect both $q_V$ and $c_V^2$.

In the small-scale limit, the contribution of the matter
fields in Eq.~({\ref{vecre}) can be neglected by assuming that 
the constants $C_i$ are background dominated for large $k$.
In this case we have
\be
\tilde{V}_i \simeq -\frac{\phi}{q_T}
 \left( 2G_{4,X}-G_{5,X}H \phi \right)\tilde{Z}_i\,.
\label{VZ}
\ee
Substituting this relation into Eq.~(\ref{Lv4}) and ignoring
the effective mass term $m_V^2 \tilde{Z}_i^2$ relative to those 
containing $(k^2/a^2)\tilde{Z}_i^2$, the second-order 
action (\ref{Lv4}) in Fourier space reduces to
\be
S_V^{(2)} \simeq \int dt\,d^3 x \sum_{i=1}^{2}\,
\frac{a^3q_V}{2}  \left( \dot{\tilde{Z}}_i^2+c_V^2\frac{k^2}{a^2}
\tilde{Z}_i^2 \right)\,.
\label{SVac}
\ee
Introducing the following quantities
\be
{\cal U}_i=z_V \tilde{Z}_i\,,\qquad
z_V=a\sqrt{q_V}\,,\qquad
\tau=\int a^{-1}dt\,,
\ee
the action (\ref{SVac}) can be expressed as
\be
S_V^{(2)} \simeq \int d\tau\,d^3 x \sum_{i=1}^{2}
\frac12 \left[ {\cal U}_i'^2+c_V^2k^2 {\cal U}_i^2
+\frac{z_V''}{z_V}{\cal U}_i^2 \right]\,,
\label{SVac2}
\ee
where a prime represents a derivative with respect to
the conformal time $\tau$.
Provided the variation of $q_V$ is not significant such that
$|\dot{q}_V| \lesssim |Hq_V|$ and $|\ddot{q}_V| \lesssim |H^2q_V|$,
we have that $c_V^2k^2 {\cal U}_i^2 \gg |(z_V''/z_V){\cal U}_i^2|$
for the perturbations deep inside the vector sound horizon
($c_V^2k^2/a^2 \gg H^2$).
Then, the equation of motion for ${\cal U}_i$ is given by
\be
{\cal U}_i''+c_V^2 k^2{\cal U}_i \simeq 0\,.
\ee
As long as the frequency $\omega_k=c_Vk$ adiabatically changes
in time, we have the following WKB solution, which is 
valid only in the regime $c_V^2k^2/a^2>H^2$ :
\be
\tilde{Z}_i=\frac{{\cal U}_i}{z_V}
\simeq \frac{1}{a\sqrt{2q_Vc_Vk}}
\left( \alpha_k e^{-ic_Vk \tau}+\beta_k e^{ic_Vk \tau}
\right)\,,
\label{Ziso}
\ee
where $\alpha_k$ and $\beta_k$ are integration constants.
Hence, for $q_V$ and $c_V$ slowly varying in time,
the perturbation $\tilde{Z}_i$ oscillates with an amplitude 
decreasing as $a^{-1}$.

For dark energy models in which the energy density of the temporal
vector component comes out at the late cosmological epoch \cite{DeFelice16}
the quantities $G_{4,X}$ and $G_{5,X}$ in Eq.~(\ref{VZ})
are usually small in the radiation and matter eras,
so the perturbation $\tilde{V}_i$ should be  suppressed.
The wave numbers $k$ relevant to the observations of large scale structures
and weak lensing correspond to $k \gg a_0H_0$ (the lower index
``0'' represents present values), so unless $c_V$ is not
much smaller than 1, the solution (\ref{Ziso}) is valid for such 
wave numbers from the vector sound horizon entry ($c_V^2k^2/a^2=H^2$) 
to today. This means that, for $q_V$ and $c_V^2$ adiabatically changing in time,
the vector perturbations $\tilde{Z}_i$ tend to be negligible with time.

\section{Scalar perturbations}
\label{scasec}

In this section we derive the equations of motion for
scalar perturbations by expanding the action (\ref{Lag})
up to quadratic order. We also introduce observables
associated with measurements of large-scale structures,
CMB, and weak lensing.

\subsection{Perturbation equations}

First of all, we define the matter
perturbation $\delta \rho_M$, as
\be
\delta \rho_M=\frac{\rho_{M,n}}{a^3} \delta J
=\frac{\rho_{M}+P_{M}}{n_0a^3} \delta J\,,
\ee
where we used Eq.~(\ref{PM}) in the second equality.
For the expansion of the matter action (\ref{Spf}) of
the scalar mode, we need to consider the perturbation
$\delta n$ of the number density $n$, as
\be
\delta n= \frac{\delta \rho_M}{\rho_{M,n}}
-\frac{{\cal N}_0^2 (\partial \chi)^2+2{\cal N}_0
\partial \chi \partial \delta j+(\partial \delta j)^2}
{2{\cal N}_0a^5}
\,,
\ee
which is expanded up to quadratic order in scalar perturbations.
Then, the second-order matter action of the scalar mode
is given by
\ba
(S_M)_S^{(2)}
&=&\int dt d^3 x \biggl\{ \frac{1}{2a^5n_0 \rho_{M,n}^2}
[ \rho_{M,n} ( \rho_{M,n}^2 \partial \delta j^2
+2a^3n_0 \rho_{M,n}^2 \partial \delta j \partial v
+2a^8n_0  \rho_{M,n}  \dot{v} \delta \rho_M
-6a^8n_0^2 \rho_{M,nn}Hv \delta \rho_M) \nonumber \\
& &\qquad \qquad -a^8 n_0 \rho_{M,nn} \delta \rho_M^2 ]
-a^3 \alpha \delta \rho_M
+\frac{\rho_{M,n}}{a^2} \partial \chi \partial \delta j
\biggr\}\,.
\label{SMS}
\ea
Varying this action with respect to $\delta j$, we obtain
\be
\partial \delta j=-a^3 n_0 \left( \partial v
+\partial \chi \right)\,.
\label{deltaj}
\ee
On account of Eq.~(\ref{deltaj}), the perturbation $\delta j$
appearing in Eq.~(\ref{SMS}) is integrated out.

We introduce the following combination
\be
\psi=\chi_V+\phi(t) \chi\,,
\ee
so that $A_i=\partial_i\psi$ for scalar perturbations. 
On using Eq.~(\ref{SMS}) with the relation (\ref{deltaj}), the
second-order action of Eq.~(\ref{Lag}) for scalar
perturbations reads
\ba
S_{S}^{(2)} & = & \int dt d^3 x\,a^{3}\,\Biggl\{-\frac{n_0 \rho_{M,n}}{2}\,\frac{(\partial v)^{2}}{a^{2}}
+\left[ n_0\rho_{M,n}\,\frac{\partial^{2}\chi}{a^2}-\dot{\delta\rho}_M
-3H\left(1+c_M^2 \right)\,\delta\rho_M \right]v
-\frac{c_M^2}{2n_0\rho_{M,n}}(\delta \rho_M)^{2}
\nonumber \\
 &  & {}-\alpha \delta\rho_M -w_{3}\,\frac{(\partial\alpha)^{2}}{a^{2}}+w_{4}\alpha^{2}
 -\left[(3Hw_{1}-2w_{4})\frac{\delta\phi}{\phi}-w_{3}\,\frac{\partial^{2}(\delta\phi)}
 {a^{2}\phi}-w_{3}\,\frac{\partial^{2}\dot{\psi}}{a^{2}\phi}
 +w_{6}\,\frac{\partial^{2}\psi}{a^{2}}\right] \alpha\nonumber \\
 &  & {}-\frac{w_{3}}{4}\,\frac{(\partial\delta\phi)^{2}}{a^{2}\phi^{2}}
 +w_{5}\,\frac{(\delta\phi)^{2}}{\phi^{2}}
 -\left[\frac{(w_{6}\phi+w_{2})\psi}{2}-\frac{w_{3}}{2}\dot{\psi}\right]
 \frac{\partial^{2}(\delta\phi)}{a^{2}\phi^{2}}\nonumber \\
 &  & {}-\frac{w_{3}}{4\phi^{2}}\,\frac{(\partial\dot{\psi})^{2}}{a^{2}}+\frac{w_{7}}{2}\,
 \frac{(\partial\psi)^{2}}{a^{2}}
 +\left(w_{1}\alpha+\frac{w_{2}\delta\phi}{\phi}\right)\frac{\partial^{2}\chi}{a^{2}}\Biggr\}\,,
\label{sscalar}
\ea
with the short-cut notations
\ba
w_{1} & = & {H}^{2} {\phi}^{3} (G_{{5,X}}+{\phi}^{2}G_{{5,{\it XX}}})
-4\,H(G_{{4}}+{\phi}^{4}G_{{4,{\it XX}}})-{\phi}^{3}G_{{3,X}}\,,
\label{w1}\\
w_{2} & = & w_1+2Hq_T\,,\label{w2}\\
w_{3} & = & -2{\phi}^{2}q_V\,,\label{w3} \\
w_{4} & = & \frac{1}{2}{H}^{3}\phi^{3}(9G_{{5,X}}-\phi^{4}G_{{5,{\it XXX}}})
-3\,H^{2} (2G_{{4}}+2\phi^{2}G_{{4,X}}+\phi^{4}G_{{4,{\it XX}}}-\phi^{6}G_{{4,{\it XXX}}}) \nonumber \\
 & &-\frac{3}{2}\,H\phi^{3}(G_{{3,X}}-\phi^{2}G_{{3,{\it XX}}})
 +\frac{1}{2}\,\phi^{4}G_{{2,{\it XX}}}\,,\label{w4} \\
w_{5} & = & w_{4}-\frac{3}{2}\,H(w_{1}+w_{2})\,, \label{w5} \\
w_{6} & = & -\phi\,\left[{H}^{2}\phi(G_{{5,X}}-{\phi}^{2}G_{{5,{\it XX}}})
-4\,H(G_{{4,X}}-{\phi}^{2}G_{{4,{\it XX}}})+\phi G_{{3,X}}\right]\,, \label{w6} \\
w_{7} & = & 2(H\phi G_{{5,X}}-2G_{{4,X}}) \dot{H}
+\left[H^{2}(G_{{5,X}}+{\phi}^{2}G_{{5,{\it XX}}})-4\,H\phi\,G_{{4,{\it XX}}}-G_{{3,X}}\right] \dot{\phi}\,.
\label{w7}
\ea
The quantity $c_M^2$ corresponds to the matter propagation speed squared 
given by
\be
c_M^2=\frac{n_0\rho_{M,nn}}{\rho_{M,n}}\,.
\label{cM}
\ee
We note that the terms containing $G_{2,F}, G_{2,Y}, g_5, G_6,G_{6,X}$
appear only in the coefficient $w_3$.
Hence, the functions $g_5(X), G_6(X)$ as well as
$G_2(F,Y)$ lead to modifications to the quadratic
action (\ref{sscalar}) through the change of $q_V$.

Varying the action $S_S^{(2)}$ with respect to
$\alpha, \chi, \delta \phi, v, \partial \psi$,
and $\delta \rho_M$, we obtain the following equations
of motion in Fourier space respectively:
\ba
& &
\delta \rho_M-2w_4 \alpha+\left( 3Hw_1-2w_4 \right)\frac{\delta \phi}{\phi}
+\frac{k^2}{a^2} \left( {\cal Y}
+w_1 \chi-w_6 \psi \right)=0\,,\label{per1} \\
& &
\left( \rho_M+P_M \right) v+
w_1 \alpha+\frac{w_2}{\phi} \delta \phi=0\,,\label{per2}\\
& &
\left( 3Hw_1-2w_4 \right)\alpha-2w_5 \frac{\delta \phi}{\phi}
+\frac{k^2}{a^2} \left[ \frac12 {\cal Y}
+w_2 \chi-\frac12 \left( \frac{w_2}{\phi}+w_6 \right) \psi
\right]=0\,,\label{per3} \\
& &
\dot{\delta \rho}_M+3H\left(1+c_M^2 \right) \delta \rho_M
+\frac{k^2}{a^2} \left( \rho_M+P_M \right)
\left( \chi+v \right)=0\,,
\label{per4}\\
& &
\dot{\cal Y}+\left( H -\frac{\dot{\phi}}{\phi} \right){\cal Y}
+2\phi \left( w_6 \alpha+w_7 \psi \right)
+\left( \frac{w_2}{\phi}+w_6 \right) \delta \phi
=0\,,\label{per5}\\
& &
\dot{v}-3Hc_M^2 v-c_M^2 \frac{\delta \rho_M}
{\rho_M+P_M}-\alpha=0\,,\label{per6}
\ea
where
\be
{\cal Y} \equiv
 \frac{w_3}{\phi}
\left( \dot{\psi}+\delta \phi+2\alpha \phi \right)\,.
\label{Ydef}
\ee
The dynamics of scalar perturbations is known by solving the
first-order differential equations (\ref{per4})-(\ref{Ydef}) for
$\delta \rho_M, {\cal Y}, v, \psi$ and the algebraic
equations (\ref{per1})-(\ref{per3}) for
$\alpha, \chi, \delta \phi$.

\subsection{Observables associated with non-relativistic matter}

A key observable related with the measurements of large-scale
structures and weak lensing is the gauge-invariant
matter density contrast $\delta$, defined by
\be
\delta \equiv \frac{\delta \rho_M}{\rho_M}+3H(1+w_M)v\,,
\ee
where $w_M \equiv P_M/\rho_M$ is the matter
equation of state.
We are interested in the evolution of non-relativistic matter perturbations (dark matter and baryons) satisfying
the conditions $w_M=0$ and $c_M^2=0$.
In this case, Eqs.~(\ref{per4}) and (\ref{per6})
reduce, respectively, to
\ba
& &\dot{\delta}-3\dot{\cal B}=-\frac{k^2}{a^2}
\left( \chi+v \right)\,,
\label{maeq1}\\
& &\dot{v}=\alpha\,,
\label{maeq2}
\ea
where ${\cal B} \equiv Hv$.

Taking the time derivative of Eq.~(\ref{maeq1}) and using Eq.~(\ref{maeq2}),
it follows that
\be
\ddot{\delta}+2H\dot{\delta}+\frac{k^2}{a^2}\Psi
=3\ddot{\cal B}+6H \dot{\cal B}\,,
\label{deleq}
\ee
where $\Psi$ is the gauge-invariant Bardeen
gravitational potential defined by \cite{Bardeen}
\be
\Psi \equiv \alpha+\dot{\chi}\,.
\ee
The growth of the matter density contrast $\delta$ is sourced
by the gravitational potential $\Psi$.
We relate $\Psi$ and $\delta$ through the modified
Poisson equation
\be
\frac{k^2}{a^2}\Psi=-4\pi G_{\rm eff} \rho_M \delta\,,
\label{Geff}
\ee
where $G_{\rm eff}$ corresponds to
the effective gravitational coupling known
by solving the perturbation Eqs.~(\ref{per1})-(\ref{Ydef})
for $\Psi$ and $\delta$. To quantify the growth rate
of $\delta$, we also define
\be
f \equiv \frac{\dot{\delta}}{H \delta}\,.
\ee
An important observable associated with RSD measurements
is the quantity $f \sigma_8$ \cite{Kaiser,Tegmark},
where $\sigma_8$ is the amplitude of over-density
at the comoving 8\,$h^{-1}$ Mpc scale
($h$ is the normalized today's Hubble parameter
$H_0=100\,h$~km\,sec$^{-1}$Mpc$^{-1}$).

Besides $\Psi$, we also introduce another gauge-invariant gravitational potential
\be
\Phi \equiv H \chi\,,
\label{Phidef}
\ee
and the gravitational slip parameter
\be
\eta \equiv -\frac{\Phi}{\Psi}\,.
\label{eta}
\ee
The effective gravitational potential associated
with the deviation of light rays in CMB and weak
lensing observations is given by \cite{AKS}
\be
\Phi_{\rm eff}=\frac12 \left( \Psi-\Phi \right)
=\frac12 \left( 1+\eta \right)\Psi\,,
\label{Phieff}
\ee
which is affected by both $\Psi$ and $\eta$.

\section{Effective gravitational coupling for matter perturbations}
\label{quasisec}

The comoving wave numbers associated with the
galaxy power spectrum for linear perturbations
are in the range
$0.01\,h$ Mpc$^{-1} \lesssim
k \lesssim 0.2\,h$ Mpc$^{-1}$ \cite{Tegmark},
which correspond to $30a_0H_0 \lesssim k \lesssim
600 a_0H_0$.
To derive analytic expressions of $G_{\rm eff}, \eta, \Psi, \Phi$
on scales relevant to the observations of large-scale structures
and weak lensing, we employ the so-called quasi-static approximation
for the perturbations inside the sound horizon.

\subsection{Quasi-static approximation on scales
deep inside the sound horizon}
\label{quasisec2}

For the theories with ${\cal L}_6=0$ and the Lagrangian ${\cal L}_2$ with no $Y$ dependence,
the no-ghost and stability conditions of scalar perturbations were
derived in Ref.~\cite{DeFelice16} in the small-scale limit.
Even for theories with $G_6 \neq 0$ and ${\cal L}_2=G_2(X,F,Y)$,
the modifications to scalar perturbations arise only through the
change of $q_V$.
In the $k \to \infty$ limit, the condition for the absence of
scalar ghosts is given by
\be
Q_{S}=\frac{a^{3}H^2q_Tq_S}{\phi^{2}(w_{1}-2w_{2})^{2}}>0\,,
\label{Qsge}
\ee
where
\be
q_S \equiv 3w_{{1}}^{2}+4q_Tw_4\,.
\label{qs}
\ee
Since the quantity $Q_S$ does not contain $w_3$, the no-ghost
condition is not modified relative to the theories studied
in Ref.~\cite{DeFelice16}.
Besides the matter propagation speed squared (\ref{cM}),
the propagation speed squared associated with
another scalar degree of freedom is given by
\be
c_{S}^{2}=\frac{\mu_c}{8H^2 \phi^2 q_Tq_Vq_S}\,,
\label{cs}
\ee
where
\ba
\mu_c &\equiv& \left[ w_6 \phi (w_1-2w_2)
+w_1 w_2 \right]^2-w_3 \left( 2w_2^2 \dot{w}_1
-w_1^2 \dot{w}_2 \right)+2w_2^2 w_3 \left(
\rho_M+P_M \right)
+\phi \left( w_1-2w_2 \right)^2 w_3 \dot{w}_6 \nonumber \\
& & +w_3(w_1-2w_2) \left[ \left( H -2\dot{\phi}/\phi \right)
w_1 w_2+\left( w_1-2w_2 \right)
\left\{ w_6 \left( H \phi -\dot{\phi} \right)+2w_7 \phi^2
\right\} \right]\,.
\label{muc}
\ea
To avoid the small-scale Laplacian instability we require
that $c_S^2>0$. This translates to $\mu_c>0$
under the three no-ghost conditions $q_T>0,q_V>0,q_S>0$.
{It should be noted that since the expression for 
$c_S^2$ contains the term $w_3$, compared to the case 
in Ref.~\cite{DeFelice16}, the new Lagrangians
${\cal L}_6$ and ${\cal L}_2=G_2(X,F,Y)$
contribute to the scalar sound speed.

In the following, we employ the quasi-static approximation
for the perturbations deep inside the sound horizon
($c_S^2k^2/a^2 \gg H^2$) \cite{Boi,DKT}.
This amounts to picking up the terms containing
$k^2/a^2$ and $\delta \rho_M$ in Eqs.~(\ref{per1})-(\ref{per6}),
see Appendix A for more detailed discussion about this approximation.
This approximation breaks down for the models in which
$c_S^2$ is very close to 0.
In the following we assume that $c_S^2$ is not very much smaller than 1,
in such a way that the condition $c_S^2k^2/a^2 \gg H^2$ holds
for the perturbations relevant to the growth of large-scale structures.
We also note that, in some dark energy models like $f(R)$
gravity \cite{fRw}, the mass $m$ of a scalar degree of freedom
can be much larger than $H$ in the past.
In our generalized Proca theories, we are interested in
the mass term $m^2 X$ in the Lagrangian ${\cal L}_2$ with $m$ 
at most of the order of $H_0$ \cite{DeFelice16}, so we can consistently ignore
its effect for discussing the perturbations deep inside the sound horizon.

In what follows, we focus on non-relativistic matter satisfying
the conditions $P_M=0$ and $c_M^2=0$.
Employing the quasi-static approximation mentioned above
for Eqs.~(\ref{per1}) and (\ref{per3}), it follows that
\ba
\delta \rho_M &\simeq&
-\frac{k^2}{a^2} \left( {\cal Y}
+w_1 \chi-w_6 \psi \right)\,,\label{per1d} \\
{\cal Y} &\simeq&
\left( \frac{w_2}{\phi}+w_6 \right)\psi
-2w_2\chi\,.
\label{per3d}
\ea
Substituting Eq.~(\ref{per3d}) into Eq.~(\ref{per1d}), we have
\be
\delta \rho_M \simeq -\frac{k^2}{a^2} \left[ (w_1-2w_2)\chi +\frac{w_2}{\phi} \psi \right]
=-\frac{k^2}{a^2} \left[ \frac{w_1-2w_2}{H}\Phi +\frac{w_2}{\phi} \psi \right]\,,
\label{quasi1}
\ee
where, in the second equality, we expressed $\chi$
in terms of $\Phi$.
{}From Eqs.~(\ref{per2}) and (\ref{per4}) we eliminate $v$ 
and obtain
\be
\dot{\delta \rho}_M+3H \delta \rho_M
+\frac{k^2}{a^2} \left( \rho_M \chi-w_1 \alpha
-\frac{w_2}{\phi} \delta \phi \right)=0\,.
\label{dotdelrho}
\ee
We take the time derivative of Eq.~(\ref{quasi1})
and eliminate the terms $\dot{\delta \rho}_M$
and $\delta \rho_M$ in Eq.~(\ref{dotdelrho}).
In doing so, we exploit Eq.~(\ref{per3d}) with
the definition of ${\cal Y}$ given in Eq.~(\ref{Ydef})
to remove the $\dot{\psi}$ term.
The perturbation $\alpha+\dot{\chi}$ can be expressed
in terms of the Bardeen gravitational potential $\Psi$.
This process leads to
\be
\phi^2 (w_1-2w_2)w_3  \Psi
+\mu_1 \Phi+\mu_2 \psi \simeq 0\,,
\label{quasi2}
\ee
where
\ba
\mu_1 &\equiv&
\frac{\phi^2}{H} \left[ \left( \dot{w}_1-2\dot{w}_2+Hw_1
-\rho_M \right)w_3-2w_2(w_2+Hw_3) \right]\,,
\label{mu1}\\
\mu_2 &\equiv&
\phi\left( w_2^2+Hw_2w_3+\dot{w}_2 w_3 \right)
+w_2 ( w_6\phi^2-w_3 \dot{\phi})\,.
\label{mu2}
\ea

We also take the time derivative of Eq.~(\ref{per3d}) and
eliminate the $\dot{\cal Y}$ and ${\cal Y}$ terms
in Eq.~(\ref{per5}).
Then, it follows that
\be
2\phi^2 w_2 \Psi+\mu_3 \Phi+\mu_4 \psi
\simeq 0\,,
\label{quasi3}
\ee
where
\ba
\hspace{-0.8cm}
\mu_3 &\equiv& \frac{2\phi}{Hw_3} \mu_2\,,
\label{mu3}\\
\hspace{-0.8cm}
\mu_4 &\equiv&
-\frac{1}{w_3} \left[ \phi^3 (w_6^2+2w_3w_7)
+\phi^2 (2w_2w_6+Hw_3w_6+w_3\dot{w}_6)
+\phi \left\{ w_2^2+Hw_2w_3+w_3(\dot{w}_2-\dot{\phi}w_6)
\right\}-2\dot{\phi}w_2w_3 \right].
\label{mu4}
\ea

We can solve Eqs.~(\ref{quasi1}), (\ref{quasi2}), (\ref{quasi3})
for $\Psi$, $\Phi$, and $\psi$, as
\ba
\Psi &\simeq&
-\frac{H(\mu_2 \mu_3-\mu_1 \mu_4)}
{\phi \mu_5}
\frac{a^2}{k^2} \rho_M \delta\,,\\
\Phi &\simeq& \frac{\phi H[2w_2 \mu_2-w_3 \mu_4(w_1-2w_2)]}
{\mu_5} \frac{a^2}{k^2} \rho_M \delta\,,\\
\psi &\simeq& \frac{\phi H[w_1w_3 \mu_3 -2w_2(\mu_1+w_3 \mu_3)]}
{\mu_5} \frac{a^2}{k^2} \rho_M \delta\,,
\ea
where
\be
\mu_5 \equiv
(w_1-2w_2) \left[ \phi (w_1-2w_2)w_3 \mu_4
-2\phi w_2 \mu_2 \right]
+Hw_2 \left[ 2w_2(\mu_1+w_3 \mu_3)
-w_1 w_3 \mu_3 \right]\,.
\label{mu5}
\ee
Note that we used the approximation
$\delta \simeq \delta \rho_M/\rho_M$, which is
valid deep inside the sound horizon.
From Eqs.~(\ref{Geff}) and (\ref{eta}),
the effective gravitational coupling and
the gravitational slip parameter are given,
respectively, by
\ba
G_{\rm eff}
&= & \frac{H(\mu_2 \mu_3-\mu_1 \mu_4)}
{4\pi \phi \mu_5}\,,
\label{Geffq}\\
\eta
&=& \frac{\phi^2[2w_2 \mu_2-w_3 \mu_4(w_1-2w_2)]}
{\mu_2 \mu_3-\mu_1 \mu_4}\,.
\label{etaq}
\ea
Under our approximation scheme, the r.h.s. of Eq.~(\ref{deleq}) is
neglected relative to the l.h.s., so that
\be
\ddot{\delta}+2H\dot{\delta}-4\pi G_{\rm eff}
\rho_M \delta \simeq 0\,,
\label{deleq2}
\ee
where we used Eq.~(\ref{Geff}).
For a given model we can integrate Eq.~(\ref{deleq}) for $\delta$
by using the analytic expression (\ref{Geffq}).
In Sec.~\ref{obsersec} we shall confirm the validity of the
above quasi-static approximation for a class of dark energy
models in generalized Proca theories.

\subsection{Estimates for $G_{\rm eff}$ and $\eta$}
\label{gravitysec}

We rewrite the effective gravitational coupling (\ref{Geffq}) and
the gravitational slip parameter (\ref{etaq}) in more convenient forms
by using physical quantities like $q_S$ and $c_S^2$ associated with
no-ghost and stability conditions (along the similar line
performed in Ref.~\cite{Tsuji15} for scalar Horndeski theories).
In doing so, we first substitute the relations
$w_1=w_2-2Hq_T$ and $w_3=-2\phi^2 q_V$ into Eq.~(\ref{Geffq})
with $\mu_{i}$ given by Eqs.~(\ref{mu1})-(\ref{mu2}),
(\ref{mu3})-(\ref{mu4}), and (\ref{mu5}).
{}From the definitions of $w_1$, $q_T$, and $w_6$
in Eqs.~(\ref{w1}), (\ref{qT}), and (\ref{w6}),
it follows that
\ba
G_{3,X} &=& -\frac{1}{2\phi^3}
\left[ w_2+w_6\phi
+8H \phi^4 G_{4,XX}-2H^2 \phi^3
(G_{5,X}+\phi^2 G_{5,XX}) \right]\,,\\
G_{4,X} &=& -\frac{1}{8H\phi^2}
\left(w_2-w_6\phi-4H^2 \phi^3 G_{5,X}\right)\,,\\
G_{4} &=& \frac{1}{8H}
\left( 4H q_T-w_2+w_6 \phi \right)\,.
\ea
On using these relations with the background
Eqs.~(\ref{be1})-(\ref{be2}), the terms $\rho_M+P_M$
and $w_7$ can be expressed as
\ba
\rho_M+P_M
&=& -2q_T \dot{H}-\frac{\dot{\phi}}{\phi}w_2\,,\\
w_7 &=& \frac{1}{2H\phi^3}
\left[ (w_2-w_6\phi )\dot{H}\phi
+(w_2+w_6\phi )H\dot{\phi} \right]\,.
\ea

We substitute these relations into Eq.~(\ref{cs})
and then express $\dot{w}_6$ with respect to $c_S^2$.
This allows us to eliminate the $\dot{w}_6$ term in the
expression of $G_{\rm eff}$ (which appears through
$\mu_4$). The resulting effective gravitational coupling
$G_{\rm eff}$ contains the time derivatives
$\dot{H}$ and $\dot{\phi}$.
Taking the time derivative of Eq.~(\ref{be3}) for the branch
$\phi \neq 0$, combining it with  Eq.~(\ref{be2}),
and eliminating the $G_2$ and $G_{2,X}$ terms
on account of Eqs.~(\ref{be1}) and (\ref{be3}),
we can write $\dot{H}$ and $\dot{\phi}$
in terms of $w_1$, $q_T$, and $w_4$.
Employing the relation (\ref{qs}) to express $w_4$
with respect to $q_S$, it follows that
\ba
\dot{H}
&=& \frac{3w_2^2-q_S}
{2q_Tq_S} (\rho_M+P_M)\,,\\
\dot{\phi}
&=&-\frac{3w_2\phi}{q_S}
 (\rho_M+P_M)\,.
\ea
After setting $P_M=0$ for non-relativistic matter,
Eqs.~(\ref{Geffq}) and (\ref{etaq}) reduce, respectively, to
\ba
G_{\rm eff}
&=& \frac{\xi_2+\xi_3}{\xi_1}\,,\label{Geff2}\\
\eta
&=&
\frac{\xi_4}{\xi_2+\xi_3}\,,\label{eta2}
\ea
with the shorthand notations
\ba
\hspace{-0.8cm}
\xi_1 &=& 4\pi \phi^2 \left( w_2+2H q_T \right)^2\,,
\label{xi1} \\
\hspace{-0.8cm}
\xi_2 &=& \left[ H\left( w_2+2Hq_T \right)-\dot{w}_1
+2\dot{w}_2+\rho_M \right]\phi^2-\frac{w_2^2}
{q_V}\,,\label{xi2}\\
\hspace{-0.8cm}
\xi_3 &=& \frac{1}{8H^2 \phi^2 q_S^3 q_T c_S^2}
\biggl[ 2\phi^2 \left\{ q_S [w_2\dot{w}_1
-(w_2-2Hq_T) \dot{w}_2]+\rho_M w_2
[3w_2(w_2+2Hq_T)-q_S] \right\} \nonumber \\
& &~~~~~~~~~~~~~~~~~~~
+\frac{q_S}{q_V}w_2
\left\{ w_2 (w_2-2Hq_T)-w_6 \phi (w_2+2Hq_T) \right\}
\biggr]^2\,,
\label{xi3} \\
\xi_4 &=& \frac{w_2+2Hq_T}{4H q_S^2q_Vq_Tc_S^2}
\biggl[ 4H^2 \phi^2 q_S^2q_Vq_Tc_S^2+
2\phi^2 q_S q_V w_2\dot{w}_2 (w_2-2Hq_T)
+w_2^2\{\phi q_S w_6(w_2+2Hq_T) \nonumber \\
& &~~~~~~~~~~~~~~~~~~~
-w_2 q_S(w_2-2Hq_T)
-2\phi^2 q_S q_V \dot{w}_1
+2\phi^2 q_V [q_S-3w_2(w_2+2Hq_T)] \rho_M \} \biggr]\,.
\label{xi4}
\ea

One can extract useful information from the expressions
(\ref{Geff2}) and (\ref{eta2}).
First, the terms proportional to $1/q_V$ in
$\xi_2$ and $\xi_3$ do not vanish for
\be
w_2=
-\phi^2 \left[\phi G_{3,X}+4H (G_{4,X}+\phi^2 G_{4,XX})
-H^2 \phi (3G_{5,X}+\phi^2 G_{5,XX}) \right]
\neq 0\,.
\ee
If the functions $G_{3,4,5}$ do not have any $X$ dependence,
which is the case for GR, then $w_2=0$ and hence $G_{\rm eff}$
is not affected by the vector contribution $q_V$.
In such cases we have $w_1=-4HG_4$ and $q_T=2G_4$
with constant $G_4$, so the quantities (\ref{xi1})-(\ref{xi4})
reduce, respectively, to
$\xi_1=64\pi G_4^2 H^2 \phi^2$,
$\xi_2=(4G_4H^2+4G_4\dot{H}+\rho_M)\phi^2$,
$\xi_3=0$, and $\xi_4=4G_4H^2 \phi^2$.
Using the relation $4G_4\dot{H}=-\rho_M$, which follows from
the background equations (\ref{be1})-(\ref{be3}),
we obtain $G_{\rm eff}=1/(16\pi G_4)$ and $\eta=1$.
Since GR corresponds to $G_4=1/(16\pi G)$, the effective
gravitational coupling reduces to $G$.

For the theories with $w_2 \neq 0$ the term $\xi_3$
does not generally vanish, so $G_{\rm eff}$ and $\eta$
generally differ from $G$ and $1$ respectively.
Under the three no-ghost and stability conditions
$q_S>0$, $q_T>0$, and $c_S^2>0$, we have that $\xi_3>0$.
Since $\xi_1$ is also positive, the presence of the term
$\xi_3/\xi_1$ in Eq.~(\ref{Geff2}) increases
the gravitational attraction.
In the expression of $\xi_2$ there exists the term
$-w_2^2/q_V$ sourced by the vector sector,
which is negative under the no-ghost condition $q_V>0$.
Hence the contribution from the vector sector
to $\xi_2/\xi_1$ works to suppress the gravitational
attraction.

In view of the recent tension between the RSD
and the Planck data \cite{Beu,Ledo,Eriksen,Vik,Planck},
we would like to discuss whether the vector
field allows the possibility for realizing the gravitational
interaction weaker than that in GR.
Since $\xi_3/\xi_1$ is positive, the {\it necessary}
condition for realizing $G_{\rm eff}$ smaller than
the Newton gravitational constant $G$
is given by $\xi_2/\xi_1<G$, i.e.,
\be
\phi^2 \left[ (w_2+2Hq_T) \left\{ H-4\pi G (w_2+2Hq_T)
\right\}-\dot{w}_1+2\dot{w}_2+\rho_M \right]
<\frac{w_2^2}{q_V}\,.
\label{weakcon}
\ee
For the function $G_2$ containing the standard Maxwell
term $F$, we may write $G_2$ of the form
$G_2=F+g_{2}(X,F,Y)$, in which case
$q_V=1+g_{2,F}+2g_{2,Y}\phi^2-4g_5H\phi
+2G_6 H^2+2G_{6,X}H^2 \phi^2$.
If the value of $q_V$ gets smaller than 1 by
the existence of functions $g_2(F,Y), g_5$, and $G_6$,
it tends to be easier to satisfy Eq.~(\ref{weakcon}).
Unlike the case of scalar-tensor Horndeski theories \cite{Tsuji15}
the condition (\ref{weakcon}) does not solely depend on quantities
associated with tensor perturbations, so
the vector field allows a more flexible possibility for
satisfying the necessary condition of weak gravity.

We would like to stress that the condition (\ref{weakcon})
is necessary but not sufficient to realize $G_{\rm eff}<G$.
Even for $\xi_2/\xi_1<G$, it can happen that the existence
of the positive term $\xi_3/\xi_1$ leads to $G_{\rm eff}$
larger than $G$. The effect of the vector sector
also appears in the expressions of $\xi_3$ and $\xi_4$.
In order to see the possibility of $G_{\rm eff}$ smaller
than $G$, we need to compute the three quantities
$\xi_1$, $\xi_2$, and $\xi_3$ for given models.
Note that, for the opposite inequality to that given
in Eq.~(\ref{weakcon}), $G_{\rm eff}$ is always larger than $G$.

\subsection{Effective gravitational coupling on the de Sitter background}
\label{dSgrasec}

On the de Sitter fixed point characterized by $\dot{\phi}=0$
and $\dot{H}=0$, it is possible to simplify
the effective gravitational coupling (\ref{Geff2}) further.
Since in this case $\dot{w}_1=\dot{w}_2=\dot{w}_6=0$, $w_7=0$,
and $\rho_M=P_M=0$, the numerator (\ref{muc}) of
$c_S^2$ reduces to
\be
\mu_c=\left[ w_2(w_2-2Hq_T)-w_6 \phi (w_2+2Hq_T)
\right] \left[  w_2(w_2-2Hq_T)-w_6 \phi (w_2+2Hq_T)
+2H\phi^2 q_V (w_2+2H q_T) \right]\,,
\label{mucd}
\ee
which is required to be positive to avoid the Laplacian instability.
Substituting Eq.~(\ref{cs}) with Eq.~(\ref{mucd})
into Eq.~(\ref{xi3}), it follows that
\be
\xi_3=\frac{w_2(w_2-2Hq_T)-w_6 \phi (w_2+2Hq_T)}
{w_2(w_2-2Hq_T)-w_6 \phi (w_2+2Hq_T)
+2H\phi^2 q_V (w_2+2H q_T)}
\frac{w_2^2}{q_V}\,.
\ee
Under the conditions $\mu_c>0$ and $q_V>0$,
the quantity $\xi_3$ is positive.
Then the effective gravitational coupling (\ref{Geff2}) reads
\be
G_{\rm eff}=\frac{H(2H\phi^2 q_V-w_6 \phi-w_2)}
{4\pi[2H\phi^2 q_V (w_2+2H q_T)+
w_2(w_2-2Hq_T)-w_6 \phi (w_2+2Hq_T)]}\,.
\label{Geffds}
\ee

In the weak-coupling limit of vector perturbations ($q_V \to \infty$),
$G_{\rm eff}$ reduces to
\be
(G_{\rm eff})_{\rm W}=
\frac{H}{4\pi (w_2+2Hq_T)}\,,
\ee
whereas, in the strong-coupling limit ($q_V \to 0$), we have
\be
(G_{\rm eff})_{\rm S}=
\frac{H(w_2+w_6 \phi)}{4\pi[w_6 \phi (w_2+2Hq_T)-
w_2(w_2-2Hq_T)]}\,.
\ee
The difference between $(G_{\rm eff})_{\rm W}$ and
$(G_{\rm eff})_{\rm S}$ is given by
\be
\Delta G_{\rm eff} \equiv
(G_{\rm eff})_{\rm W}-(G_{\rm eff})_{\rm S}
=\frac{Hw_2^2}{2\pi[(w_2+2Hq_T)
\left\{ w_2(w_2-2Hq_T)-w_6 \phi (w_2+2Hq_T)
\right\}]}\,.
\ee
If the condition
\be
(w_2+2Hq_T)
\left\{ w_2(w_2-2Hq_T)-w_6 \phi (w_2+2Hq_T)
\right\}>0
\label{Geffcon}
\ee
is satisfied, it follows that $(G_{\rm eff})_{\rm W}>(G_{\rm eff})_{\rm S}$.
In this case, the effective gravitational coupling tends to decrease
for a stronger coupling of vector modes (i.e., for smaller $q_V$).
In Sec.~\ref{obsersec} we shall consider a class of generalized
Proca theories to see how the change of $q_V$ modifies $G_{\rm eff}$.

\section{Observables in a concrete dark energy model}
\label{obsersec}

We study the evolution of perturbations associated with the
observations of large-scale structures, weak lensing, and CMB
for a dark energy model in generalized Proca theories.
Let us consider theories given by the functions
\ba
& &
G_2(X,Y,F)=b_2 X^{p_2}
+\left[ 1+g_2(X) \right]F\,,\qquad
G_3(X)=b_3X^{p_3}\,,\qquad
G_4(X)=\frac{1}{16\pi G}+b_4X^{p_4}\,,\nonumber \\
& &
G_5(X)=b_5X^{p_5}\,,\qquad
g_5(X)=\tilde{b}_5 X^{q_5}\,,\qquad
G_6(X)=b_6X^{p_6}\,,
\label{G23456}
\ea
where $G$ is the Newton gravitational constant,
$b_{2,3,4,5,6}$, $\tilde{b}_{5}$, $p_{2,3,4,5,6}$,
$q_{5}$ are constants, and $g_2(X)$ is an arbitrary
function of $X$ with $g_2(0)=0$. Compared to the model studied in
Ref.~\cite{DeFelice16}, there exists additional functional freedoms in 
$g_2(X)$ (not necessarily proportional to $X^{p_4-1}$), 
$\tilde{b}_5 X^{q_5}$ (not necessarily satisfying $q_5=p_5-1$), 
and $b_6X^{p_6}$.
The quantity $F$ vanishes on the FLRW background, so
they do not affect the background equations of motion. 
Since the background has $Y=0$, by adopting the Taylor 
expansion around $Y=0$, it is sensible to include a further 
additional term of the form $\tilde{b}_2YX^{q_2}$ in $G_2$. 
However, for simplicity, we do not consider this term in the following. 
Since the background has a non-vanishing $X$, we 
do not adopt the Taylor expansion with respect to $X$. 

\subsection{Cosmological background}

For the powers $p_{3,4,5}$ given by
\be
p_3=\frac12 \left( p+2p_2-1 \right)\,,\qquad
p_4=p+p_2\,,\qquad
p_5=\frac12 \left( 3p+2p_2-1 \right)\,,
\label{p345}
\ee
the background solution of the form
\be
\phi^{p} \propto H^{-1}
\label{phiH}
\ee
can be realized \cite{DeFelice16}, where $p$ is a positive constant.
The vector Galileon \cite{Heisenberg} corresponds to
the powers $p=p_2=1$.
For positive $p$ the temporal vector component
$\phi$ is small in the early cosmological epoch, but
it grows with the decrease of $H$ to give rise to
the late-time cosmic acceleration.
According to the stability analysis around a late-time
de Sitter fixed point, it is always a stable attractor \cite{DeFelice16}.

Since we are interested in the cosmological evolution after
the end of the radiation era, we take into account non-relativistic
matter alone for the matter Lagrangian ${\cal L}_M$ (unlike
Ref.~\cite{DeFelice16} in which radiation is also present).
We introduce the matter density parameter
$\Omega_m=8\pi G\rho_M/(3H^2)$
and the dimensionless quantities
\be
y \equiv \frac{8\pi G\,b_2\phi^{2p_2}}
{3H^2\,2^{p_2}}\,,\qquad
\beta_i \equiv
\frac{p_ib_i}{2^{p_i-p_2}p_2b_2}
\left( \phi^p H \right)^{i-2}\,,
\label{ydef}
\ee
where $i=3,4,5$ and $\beta_i$'s are constants from Eq.~(\ref{phiH}).
For the branch $\phi \neq 0$ of Eq.~(\ref{be3}), we have the
following relation
\be
1+3\beta_3+6(2p+2p_2-1)\beta_4
-(3p+2p_2)\beta_5=0\,,
\ee
which can be used to express $\beta_3$ in terms
of $\beta_4$ and $\beta_5$.

The dark energy density parameter is given by
\be
\Omega_{\rm DE}=1-\Omega_m=
\frac{6p_2^2(2p+2p_2-1)\beta_4-p_2(p+p_2)
(1+4p_2 \beta_5)}{p_2(p+p_2)}y\,,
\ee
which satisfies the differential equation
\be
\frac{d\Omega_{\rm DE}}{d{\cal N}}
=\frac{3(1+s)\Omega_{\rm DE} (1-\Omega_{\rm DE})}
{1+s\Omega_{\rm DE}}\,,
\ee
where ${\cal N}=\ln a$ and $s=p_2/p$.
{}From the matter-dominated fixed point characterized by
$\Omega_{\rm DE}=0$, the solutions finally approach
a de Sitter attractor with $\Omega_{\rm DE}=1$.
The equation of state of dark energy depends on
$\Omega_{\rm DE}$, as
\be
w_{\rm DE}=-\frac{1+s}{1+s\Omega_{\rm DE}}\,,
\ee
which evolves from $-1-s$ (matter era) to
$-1$ (de Sitter epoch).
The likelihood analysis based on the SN Ia, CMB, and
BAO data showed that the constant $s$ is constrained
to be $0 \le s<0.36$ \cite{DT12}.

\subsection{Evolution of perturbations}

The theoretical consistency of the model (\ref{G23456}) requires
that the six quantities $q_T,c_T^2,q_V,c_V^2,q_S,c_S^2$
are positive in the small-scale limit.
In Ref.~\cite{DeFelice16} the parameter space consistent
with these conditions was discussed for the specific
functions $b_6=0$, $g_2=-2c_2G_{4,X}$, and $g_5=d_2G_{5,X}/2$.
The generalization to the model (\ref{G23456}) modifies
neither the background equations of motion nor the second-order
action of tensor perturbations, but the evolution of vector
perturbations is subject to change. The scalar perturbation
is also affected by the new terms of intrinsic vector modes
through the change of $q_V$.
In the following, we investigate how the new terms affect
the evolution of scalar perturbations and observable quantities.
The evolution of vector perturbations is discussed 
at the end of this section.

For the model given by the functions (\ref{G23456}),
the parameter $q_V$ reads
\be
q_V=1+g_2-4\tilde{b}_5 X^{q_5}H\phi
+2b_6(1+2p_6)H^2 X^{p_6}\,,
\label{qVex}
\ee
where the last term arises for the theories with
$b_6 \neq 0$.
{}From Eq.~(\ref{phiH}) the last term of Eq.~(\ref{qVex})
is proportional to $\phi^{2(p_6-p)}$, so it is constant for $p_6=p$.
Depending on the sign of the term $2b_6(1+2p_6)$,
$q_V$ is either larger or smaller than 1.

The variables $w_2$ and $q_T$ can be expressed
in the following forms
\ba
w_2
&=&-\frac{2^{1-p_2/2}}{\sqrt{24\pi G}}p_2
\phi^{p_2} \sqrt{b_2 y}
\left[ 1+6\beta_4 (1-2p_2-2p)+2\beta_5 (3p+2p_2)
\right]\,,\\
q_T
&=& \frac{1}{8\pi G} \left[ 1+6\beta_4 p_2 \left( \frac{1}{p_2+p}
-2 \right)y+6\beta_5 p_2y \right]\,.
\ea
In the asymptotic past where $\Omega_{\rm DE}$ is negligibly small,
we have $y \to 0$ and hence $w_2 \to 0$ and $q_T \to 1/(8\pi G)$.
This means that, in the early matter era, the quantities
$\xi_i$'s in Eq.~(\ref{xi1})-(\ref{xi4}) are approximately given by
$\xi_1 \simeq H^2\phi^2/(4\pi G^2)$,
$\xi_2 \simeq H^2\phi^2/(4\pi G)$,
$\xi_3 \simeq 0$, and
$\xi_4 \simeq H^2 \phi^2/(4\pi G)$,
respectively,
where we used the approximate background equation of motion
$\dot{H} \simeq -4\pi G \rho_M$ (neglecting the contribution
of dark energy density).
Then, in the early matter-dominated epoch,
the effective gravitational coupling (\ref{Geff2})
and the slip parameter (\ref{eta2}) are close to
$G$ and $1$, respectively.

After the dark energy dominance the quantity $w_2$ starts to
be away from 0, which leads to the deviation of $G_{\rm eff}$
from $G$. {}From Eq.~(\ref{Geffds}) the effective gravitational
coupling on the de Sitter solution is given by
\be
\frac{G_{\rm eff}}{G}
=\frac{(p+p_2)[q_V u^2-2p_2y\{1-6\beta_4(2p+2p_2-3)
+2\beta_5(3p+2p_2-3) \}]}{{\cal F}_G}\,,
\label{Geffdsmo}
\ee
where $u=\sqrt{8\pi G}\phi$, and
\ba
{\cal F}_G
&=&q_Vu^2 \left[ p+p_2+6\beta_4 p_2y+p_2(p+p_2)
\{ 1-6\beta_4 (1+2p+2p_2)+2\beta_5 (3+3p+2p_2) \} y \right]
\nonumber \\
&&+ 2p_2y [ (p+p_2) \{ -1+6\beta_4 (2p+2p_2-3)
+\beta_5 (6-6p-4p_2) \}+6p_2 \{ 18\beta_4^2(2p+2p_2-1) \nonumber \\
& &-\beta_4 [1+\beta_5(30p+28p_2-6)] +6\beta_5^2(p+p_2)\}y]\,.
\ea
The value of $G_{\rm eff}$ at the de Sitter attractor depends on
the parameters $p, p_2,\beta_4,\beta_5$ and the quantities
$y,q_Vu^2$. Let us consider the constant $q_V$ model
realized by the non-vanishing Lagrangian ${\cal L}_6$ with
\be
p_6=p\,,\qquad g_2=0\,,\qquad \tilde{b}_5=0\,.
\label{pp6}
\ee
%

%%%%%%%%%%%%%%%%%%%%%%%%%%%%%%
\begin{figure}
\begin{center}
\includegraphics[height=3.6in,width=3.7in]{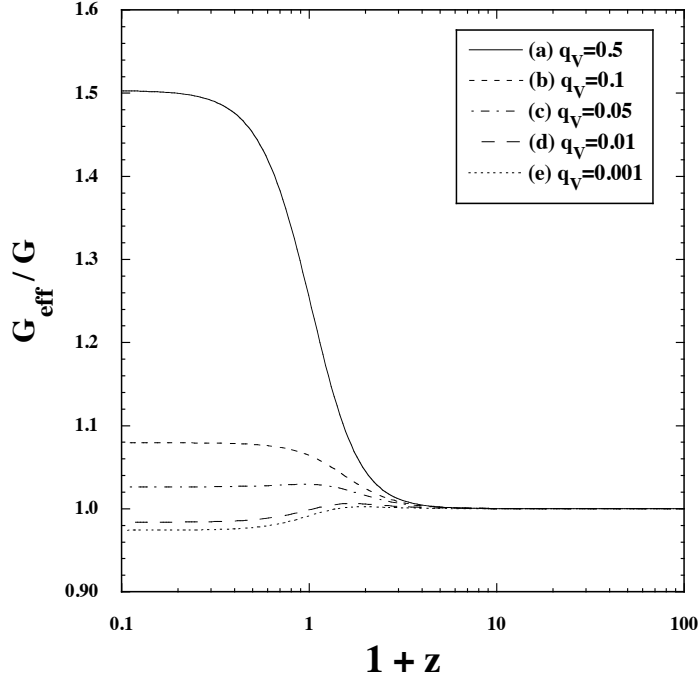}
\end{center}
\caption{\label{fig1}
Evolution of $G_{\rm eff}/G$ for the model parameters
$p_2=1/2$, $p=p_6=5/2$, $g_2=0$,
$\tilde{b}_5=0$, $\beta_4=10^{-4}$, $\beta_5=0.052$,
$\lambda=1$ with $q_V=0.5,0.1,0.05,0.01,0.001$
(from top to bottom). The present epoch (the redshift $z=0$)
is identified as $\Omega_{\rm DE}=0.68$.}
\end{figure}
%%%%%%%%%%%%%%%%%%%%%%%%%%%%%%

In Fig.~\ref{fig1} we plot the evolution of $G_{\rm eff}/G$ for
$p_2=1/2$, $p=p_6=5/2$, $\beta_4=10^{-4}$, and $\beta_5=0.052$
versus the redshift $z=a_0/a-1$ with five different values of $q_V$.
We choose the negative coefficient $b_2=-m^2 (8\pi G)^{p_2-1}$
(where $m^2>0$) with $\lambda \equiv u^{p}H/m$.
For the model parameters given above,
all the no-ghost and stability conditions of tensor, vector, and
scalar perturbations are consistently satisfied.
In Fig.~\ref{fig1} we see that $G_{\rm eff}$ is close to $G$
in the early matter era independent of $q_V$,
but the late-time evolution of $G_{\rm eff}$ is different
depending on the values of $q_V$.

For the model parameters used in Fig.~\ref{fig1},
the asymptotic values of $y$ and $u$
at the de Sitter attractor are given, respectively,
by $y=-0.906$ and $u=1.252$.
The analytic estimation (\ref{Geffdsmo}) shows that, for smaller
$q_V$, the effective gravitational coupling at the de Sitter
fixed point decreases, e.g., $G_{\rm eff}/G=1.503$ for $q_V=0.5$
and $G_{\rm eff}/G=0.974$ for $q_V=0.001$.
In fact, we have numerically confirmed that the condition
(\ref{Geffcon}) is satisfied for the model parameters
used in Fig.~\ref{fig1}.
Thus, for $q_V$ close to 0, it is possible to realize
$G_{\rm eff}$ smaller than $G$.

We also numerically computed the quantities $\xi_1,\xi_2,\xi_3$ and
found that the contribution $\xi_2/\xi_1$ to $G_{\rm eff}$ becomes
negative at low redshifts in the numerical simulation of Fig.~\ref{fig1}.
This is overwhelmed by the positive contribution $\xi_3/\xi_1$
to $G_{\rm eff}$, such that $G_{\rm eff}$ stays positive.
Thus, the necessary condition (\ref{weakcon}) for realizing
$G_{\rm eff}<G$ is satisfied for all the cases shown in Fig.~\ref{fig1}, but
we need to evaluate the $\xi_3/\xi_1$ term for each value of $q_V$
to discuss whether weak gravity is really possible.

In the left panel of Fig.~\ref{fig2}, we show the evolution of $f\sigma_8$
for several different values of $q_V$ derived by numerically integrating
the perturbation Eqs.~(\ref{per1})-(\ref{per6}).
We choose the comoving wave number $k=230a_0H_0$,
which is within the linear regime of perturbations in the
observations of large-scale structures \cite{Tegmark}.
We recall that, under the quasi-static approximation on scales
deep inside the sound horizon ($c_S^2k^2/a^2 \gg H^2$),
the matter perturbation obeys Eq.~(\ref{deleq2}) with
$G_{\rm eff}$ given by Eq.~(\ref{Geff2}).
In the numerical simulations of Fig.~\ref{fig2} the sound speed
squared tends to be larger for smaller $q_V$, whose present value is
in the range ${\cal O}(0.1)<c_S^2<{\cal O}(10^2)$.
By solving Eq.~(\ref{deleq2}) with~(\ref{Geff2}) numerically, we confirmed that
the evolution of $\delta$ obtained under the quasi-approximation exhibits 
very good agreement with
the full numerical solutions of Eqs.~(\ref{per1})--(\ref{per6}).
In fact, the theoretical curves of $f\sigma_8$ derived under the quasi-static 
approximation for the modes $c_S^2k^2/a^2 \gg H^2$
are almost indistinguishable from those obtained by full integrations.

%%%%%%%%%%%%%%%%%%%%%%%%%%%%%%
\begin{figure}
\begin{center}
\includegraphics[height=3.4in,width=3.5in]{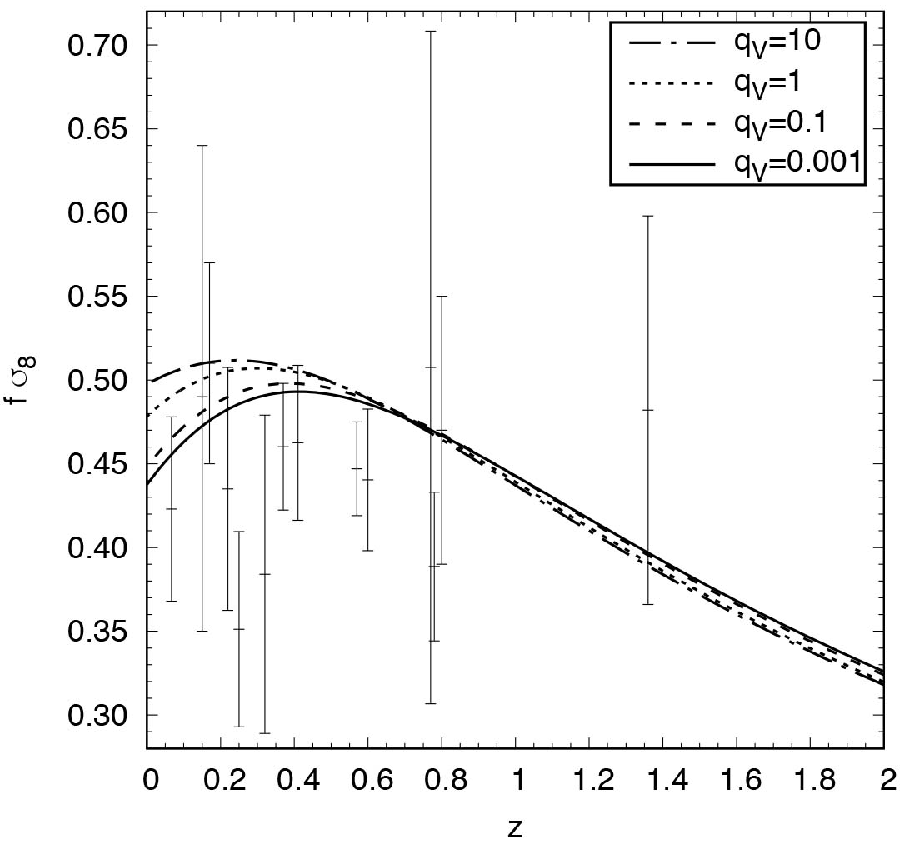}
\includegraphics[height=3.4in,width=3.5in]{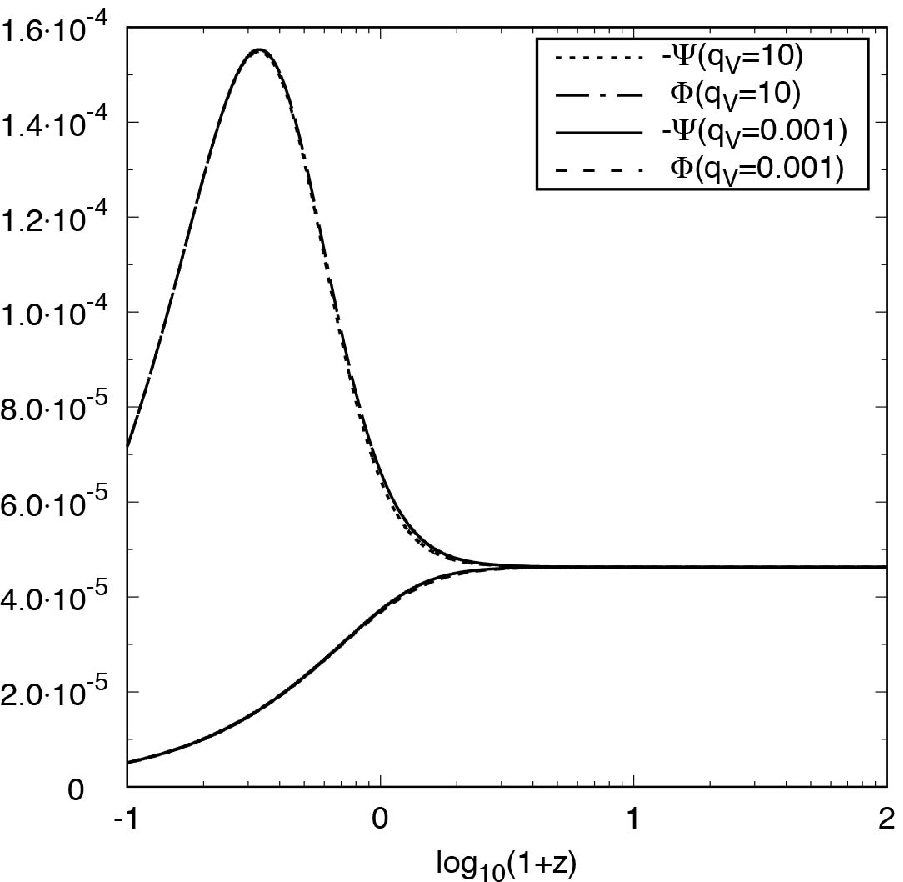}
\end{center}
\caption{\label{fig2}
(Left) Evolution of $f\sigma_8$  for the same
model parameters as those used in Fig.~\ref{fig1} with
$q_V=10,1,0.1,0.001$. The initial conditions of perturbations
are chosen to match those under the sub-horizon approximation
discussed in Sec.~\ref{quasisec} with the comoving wave number
$k=230 a_0H_0$ and $\sigma_8(z=0)=0.82$.
The black points with error bars correspond to the bounds of
$f\sigma_8$ constrained from the data of
redshift-space-distortion
measurements \cite{2dF,6dF,Wiggle,SDSS,BOSS,Torre,Okumura}.
(Right) Evolution of the gravitational potentials $-\Psi, \Phi$
for $q_V=10,0.001$.}
\end{figure}
%%%%%%%%%%%%%%%%%%%%%%%%%%%%%%

As we see in Fig.~\ref{fig2}, the theoretical values
of $f\sigma_8$ in low redshifts get smaller for decreasing $q_V$.
This behavior reflects the fact that $G_{\rm eff}$ at the de Sitter
fixed point tends to be smaller for $q_V$ closer to 0.
In Fig.~\ref{fig2} we also show the observational data constrained from
the RSD measurements (including the recent FastSound data \cite{Okumura}
measured at the highest redshift $z=1.4$).
To plot the theoretical curves, we have chosen the value
$\sigma_8(z=0)=0.82$ constrained by the recent Planck CMB data \cite{Planck}.
The theoretical prediction is in tension with some of the RSD data,
but this property also persists in the $\Lambda$CDM model for
$\sigma_8(z=0)$ constrained from Planck observations.
The tension reduces for smaller $\sigma_8(z=0)$
constrained from the WMAP data \cite{WMAP}.
In any case, the present RSD data are not sufficiently
accurate to place tight constraints on model parameters of the theory.
It is however interesting to note that the models with different values of
$q_V$ can be potentially distinguished from each other in future
RSD measurements.

In the right panel of Fig.~\ref{fig2} we also plot
the evolution of the gravitational potentials for $q_V=10, 0.001$.
As in the case of GR, both $-\Psi$ and $\Phi$ stay nearly constant
in the deep matter era with the slip parameter $\eta$ very close to 1.
They start to vary around the end of the matter-dominated epoch,
but the difference between $-\Psi$ and $\Phi$ is small.
Hence the evolution of the effective gravitational potential
$-\Phi_{\rm eff}$ defined by Eq.~(\ref{Phieff}) is similar to that of $-\Psi$ and $\Phi$.
The deviation of the slip parameter $\eta$ from 1 is typically
insignificant for theoretically consistent model parameters.

In Fig.~\ref{fig2} we see that the gravitational potentials
are enhanced for $q_V=10$ after the onset
of cosmic acceleration. This enhancement occurs due to the
strong gravitational coupling with $G_{\rm eff}>G$.
On the other hand, for $q_V=0.001$, both $-\Psi$ and $\Phi$
start to decay after the end of the matter era.
Thus, it should be possible to distinguish the models with large and small
values of $q_V$ from the integrated Sachs-Wolfe effect
of CMB observations.

\begin{figure}
\begin{center}
\includegraphics[width=3.4in]{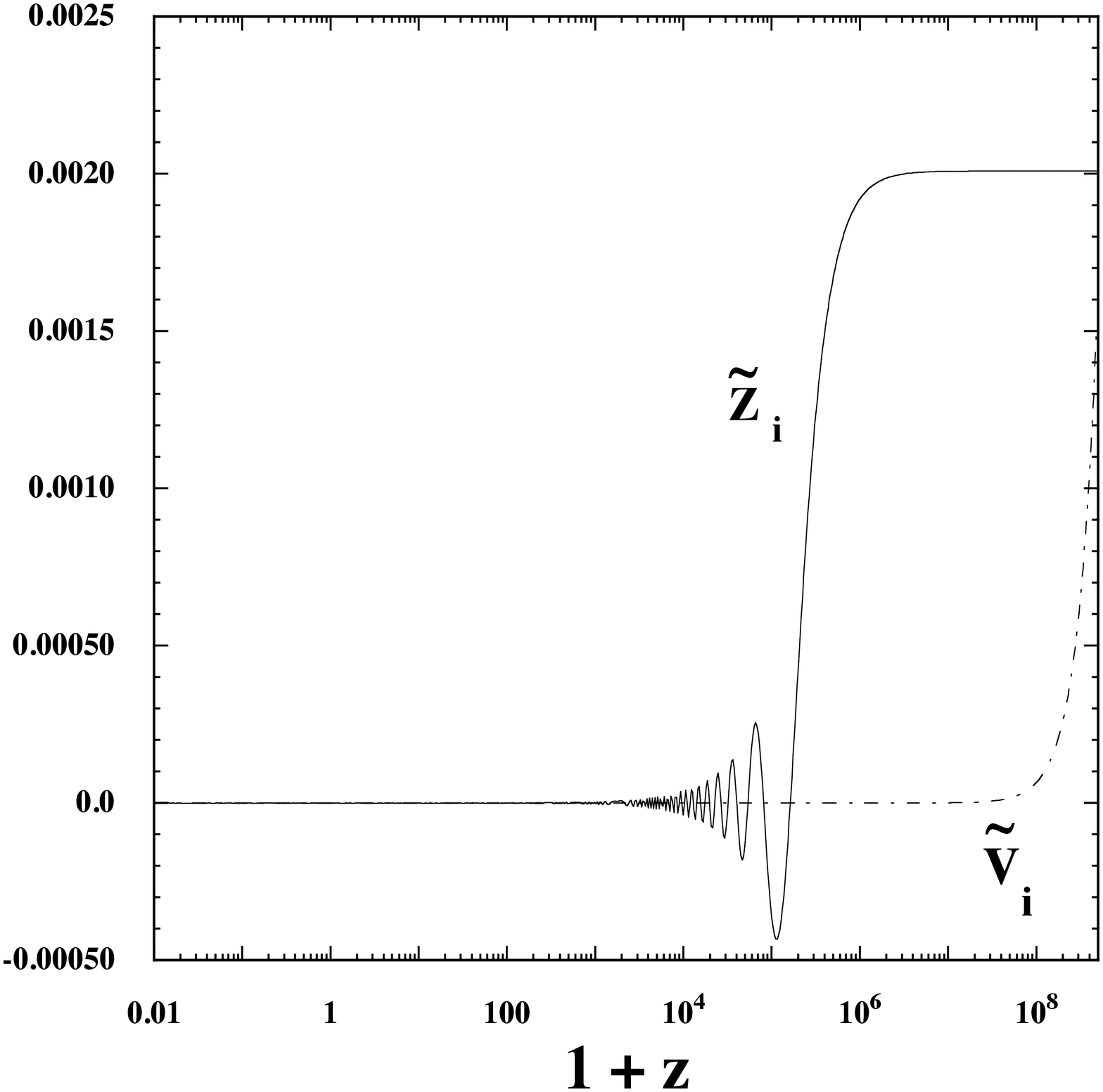}
\includegraphics[width=3.6in]{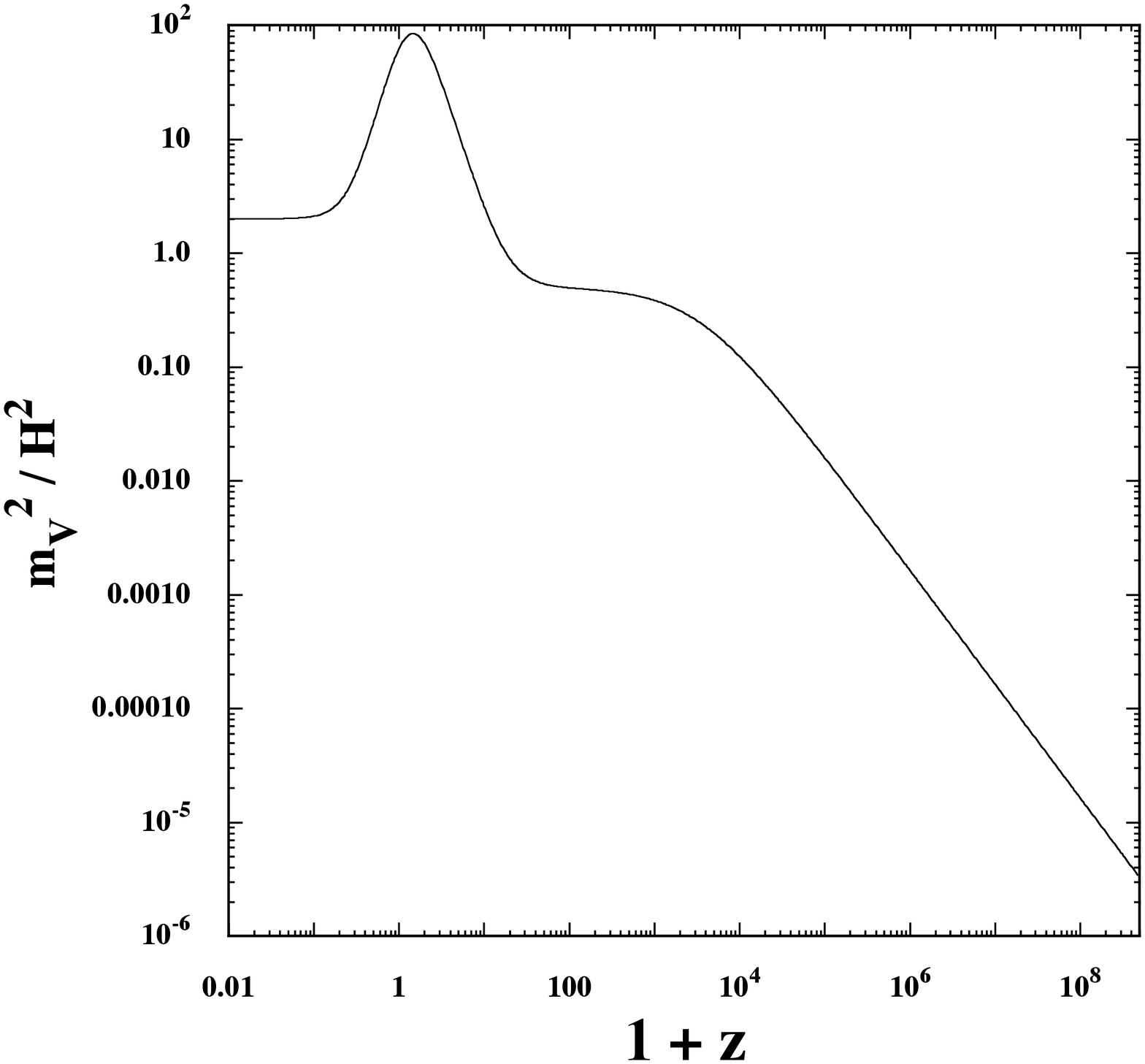}
\end{center}
\caption{\label{fig3}
Evolution of the vector perturbations $\tilde{Z}_i$ (normalized 
by $1/\sqrt{8\pi G}$) and $\tilde{V}_i$ for the case $q_V=0.001$ (left) 
and evolution of the vector mass squared  
$m_V^2$ divided by $H^2$ (right).
The model parameters are the same as those used in 
Fig.~\ref{fig1} with $q_V=0.001$.
The initial conditions are chosen, at the redshift 
$z=4.77 \times 10^{8}$, as
$\Omega_{\rm DE}=9.74\times10^{-38}$, 
$\Omega_r\equiv8\pi G\rho_r/(3H^2)=1-6.888\times10^{-6}$, 
$\tilde{Z}_i=2.0095\times10^{-3}/\sqrt{8\pi G}$,
$d\tilde{Z}_i/d{\cal N}=-10^{-8}/\sqrt{8\pi G}$, and 
$\tilde{V}_i=0.0015$ with $v_{i,r}=v_{i,m}$.
We choose the comoving wave number to be 
$k=230 a_0H_0$.}
\end{figure}

Finally, we discuss the evolution of vector perturbations from
the deep radiation era to the de Sitter epoch. 
For this purpose, we take into account radiation besides 
non-relativistic matter in the forms
$\rho_M=\rho_r+\rho_m$ and $P_M=\rho_r/3$ with 
the velocity perturbations $v_{i,r}$, $v_{i,m}$ and 
solve Eqs.~(\ref{vecre}) and (\ref{Zieq}) numerically.
In the left panel of Fig.~\ref{fig3}, the evolution of 
$\tilde{Z}_i$ and $\tilde{V}_i$ 
is plotted for $q_V=0.001$ and the wave number $k=230a_0H_0$.
At the initial stage of the radiation
era the perturbations are outside the vector sound horizon 
($c_V^2k^2/a^2<H^2$), in which regime the dynamical field 
$\tilde{Z}_i$ is nearly frozen. 
After the entry of the vector sound horizon, $\tilde{Z}_i$ 
starts to oscillate with a decreasing amplitude. 
In this regime, the evolution of $\tilde{Z}_i$ is well described 
by the WKB solution given by Eq.~(\ref{Ziso}). 
As we see in Fig.~\ref{fig3}, 
the perturbation $\tilde{V}_i$ does not grow either.

In the right panel of Fig.~\ref{fig3}, we show the evolution 
of the mass squared $m_V^2={\cal C}_2/q_V$ of 
the dynamical vector field $\tilde{Z}_i$.
The ratio $m_V^2/H^2$ grows from the radiation era to today 
and it finally approaches the asymptotic value $m_V^2/H^2=2$
at the de Sitter attractor. For small $q_V$ closer to 0, 
there is a tendency that the mass $m_V$ gets larger
than the order of $H$ at low redshifts. 
In such cases the oscillations of $\tilde{Z}_i$ are also 
present even for small $k$, but the
amplitude of $\tilde{Z}_i$ does not increase. 
In summary, there is no growth of $\tilde{Z}_i$ 
for the dark energy model studied above.

\section{Conclusions}
\label{consec}

One promising way to tackle dark energy and cosmological constant problems is to invoke new dynamical degrees of freedom in addition to those appearing in the standard model of particle physics. Modifications in form of an additional scalar degree of freedom have been mostly studied in the literature.
Among them the Galileon and Horndeski interactions
received much attention, as the latter being the most general
scalar-tensor theories with second-order equations of motion.
On the other hand, the presence of a vector degree of freedom can also induce interesting phenomenology besides providing a self-acceleration of the Universe.

In this work, we followed this latter approach and considered the most general vector-tensor interactions in form of generalized Proca theories with five propagating degrees of freedom, i.e., the two tensor gravitational degrees of freedom and the two transverse and one longitudinal mode of the vector field. To realize some non-trivial cosmological dynamics with a
gauge-invariant vector field, one usually needs to introduce spatial components of it at the background level.
In our case the $U(1)$ gauge symmetry
is explicitly broken, so that the existence of the temporal vector component can lead to interesting cosmological solutions with a late-time de Sitter attractor.

The action of our generalized Proca theories has been  constructed in such a way that time derivatives higher
than second order do not arise to avoid the Ostrogradski
instability. The temporal component $\phi$ of the vector field, which appears as an auxiliary field, can be entirely expressed in term of the Hubble expansion rate $H$.
The de Sitter solutions, which are relevant to dark energy,
can be realized for constant values of $\phi$ and $H$.
We obtained second-order actions
of tensor, vector, and scalar perturbations on top of
the general FLRW background in the presence of a matter fluid. 
This allowed us to derive general conditions for avoiding ghosts and Laplacian instabilities in the small-scale limit.

In difference to the previous analysis, the perturbations coming from the 
sixth-order Lagrangian  ${\cal L}_6$ and the quadratic Lagrangian ${\cal L}_2$
containing the $X,F,Y$ dependence (which preserves the parity invariance)
are included as well. The presence of purely vector interactions in ${\cal L}_2, g_5, {\cal L}_6$ has important impact on
the no-ghost and stability conditions for vector perturbations
and on the sound speed of scalar perturbations.
To guarantee the absence of any theoretical pathology,
we require that six no-ghost and stability conditions
are satisfied. This permits to shrink the allowed parameter
space of the theory drastically.

The main goal of this work was to study observational signatures
of generalized Proca theories related with linear cosmological perturbations.
For this purpose, we derived the full perturbation equations of motion
for tensor, vector, and scalar modes and then analytically
obtained the effective gravitational coupling
$G_{\rm eff}$ with matter density perturbations and the slip parameter
$\eta$ by employing the quasi-static approximation on scales
deep inside the sound horizon.
In view of the recent tension between the data of
redshift-space distortions and CMB, we identified the necessary condition for realizing $G_{\rm eff}$ smaller than the Newton gravitational constant $G$.
One can nicely observe the important impact of intrinsic vector modes
on $G_{\rm eff}$ in the quantity $q_V$ associated with the vector
no-ghost condition. For smaller $q_V$ there is a tendency that
$G_{\rm eff}$ decreases, so the vector field plays an important role
to modify the gravitational interaction on cosmological scales relevant to
the observations of large-scale structures and weak lensing.

For concreteness, we have considered a class of dark energy models
in which the temporal vector component $\phi$ is of the form
$\phi^{p} \propto H^{-1}$ with $p>0$.
This solution, which has a late-time de Sitter attractor, can be
realized for the functions $G_{2,3,4,5,6}$ given by Eq.~(\ref{G23456})
with the powers (\ref{p345}).
As we see in Fig.~\ref{fig1}, it is indeed possible to realize $G_{\rm eff}<G$
for small $q_V$, while satisfying six no-ghost and stability conditions.
We also numerically integrated the scalar perturbation equations of motion
to study the evolution of the growth rate $f\sigma_8$ as well as
the gravitational potentials $\Psi$ and $\Phi$.
We confirmed that the full numerical results show excellent agreement
with those derived under the quasi-static approximation for the
perturbations deep inside the sound horizon.
As we see in Fig.~\ref{fig2}, the evolution of observables is quite
different at low redshifts depending on the values of $q_V$.
Since the dark energy equation of state $w_{\rm DE}$ is also smaller
than $-1$, it is possible to distinguish our model from the
$\Lambda$CDM model according to both expansion history
and cosmic growth.

Concerning the vector perturbations, we have also provided analytic estimation for
the evolution of the transverse vector modes. This analytic estimation has been also 
confirmed by numerically solving the perturbations equations (\ref{vecre}) 
and (\ref{Zieq}) for the model (\ref{G23456}).
The evolution of the vector modes is characterized as follows: 
far outside the vector sound horizon,
the perturbations $\tilde{Z}_i$ are 
nearly constants. After the horizon entry ($c_V^2k^2/a^2>H^2$),
the perturbations start to decay with oscillations. 
Thus, there is no growth for the dynamical vector 
fields $\tilde{Z}_i$.

We have thus shown that generalized Proca theories offer a nice possibility
for realizing a dark energy model with peculiar observational signatures.
It is of interest to put observational constraints on the allowed parameter
space of the model, which we leave for a future work.

\section*{Acknowledgements}

We would like to thank M.\ Motta and F.\ Piazza for very useful
discussions. We are also grateful to T. Okumura for providing us with
the recent RSD data. ADF was supported by JSPS KAKENHI Grant Numbers
16K05348, 16H01099.  LH acknowledges financial support from Dr.~Max
R\"ossler, the Walter Haefner Foundation and the ETH Zurich
Foundation.  RK is supported by the Grant-in-Aid for Research Activity
Start-up of the JSPS No.\,15H06635. The work of SM was supported in 
part by JSPS KAKENHI Grant Number 24540256 and World Premier 
International Research Center Initiative (WPI), MEXT, Japan. 
ST is supported by the Grant-in-Aid for Scientific Research Fund of 
the JSPS Nos.~24540286, 16K05359, and MEXT KAKENHI Grant-in-Aid for 
Scientific Research on Innovative Areas ``Cosmic Acceleration'' 
(No.\,15H05890). YZ is supported by the Strategic Priority Research 
Program ``The Emergence of Cosmological Structures'' of the Chinese 
Academy of Sciences, Grant No. XDB09000000.

%%%%%%%%%
\appendix
%%%%%%%%%

\section{Sub-horizon limit and quasi-static approximation}

In this Appendix we shall clarify the distinction between the sub-horizon limit 
and the quasi-static approximation.

In the sub-horizon approximation we suppose that modes of interest have physical momenta $k/a$ sufficiently higher than the Hubble expansion rate $H$ (but sufficiently lower than the cutoff of the theory under consideration). Let us then introduce a small bookkeeping parameter $\epsilon$ ($\ll 1$) so that $Ha/k={\cal O}(\epsilon)$. In the sub-horizon limit ($\epsilon\ll 1$) it makes perfect sense to consider a dispersion relation for each propagating mode since there is a clear separation between the scale of the background and that of the perturbation. Assuming that the modes of interest approximately have linear dispersion relations in the sub-horizon limit, it is easy to see that a time derivative acted on perturbation variables is of order $H\times {\cal O}(\epsilon^{-1})$. With this assignment, we keep the lowest-order part of the quadratic action written in terms of canonically normalized perturbation variables (after eliminating non-dynamical variables of course). We can consider this procedure as the sub-horizon limit in the context of cosmological perturbations.

In the scalar perturbation sector of the system considered in the present paper, there are two propagating degrees of freedom, one from gravity and the other from dust matter. They follow coupled second-order differential equations. Therefore, a general solution in the scalar sector is a linear combination of four independent modes. 
This means that we can derive a fourth-order differential equation for 
one master variable, e.g., the gauge-invariant density contrast. 
We can also express all the other (dynamical and non-dynamical) variables, e.g., 
two gauge-invariant potentials, as linear combinations of the master variable 
and its derivatives up to third order.

When the sound speed of the degree of freedom from gravity is of order unity, one can easily 
show that the scalar sector includes two fast modes and two slow modes since the 
sound speed of dust matter is zero. The two fast modes have the time scale of order $a/k$, 
while the two slow modes have the time scale of order $1/H$ ($\gg a/k$).

The quasi-static approximation is nothing but dropping the fast modes and keeping the slow modes in order to describe an adiabatic evolution of the system. In practice we can easily take the quasi-static approximation: we start with the fourth-order differential equation for one master variable and consider that a time derivative acted on the master variable is of order $H\times {\cal O}(\epsilon^0)$ (instead of $H\times {\cal O}(\epsilon^{-1})$). By keeping the leading order contribution in the small $\epsilon$ limit with this new assignment, one obtains a second-order differential equation for the master variable. This is the equation of motion in the quasi-static approximation describing the two slow modes only. From the equation of motion for the gauge-invariant density contrast in the quasi-static approximation, one can easily read off the effective gravitational constant $G_{\rm eff}$. One can also apply the quasi-static approximation to the expressions of the two gauge-invariant potentials to obtain the Poisson equation and the slip parameter $\eta$. The expressions for $G_{\rm eff}$ and $\eta$ obtained in this way completely agree with those obtained in the main text by a different method.

%%%%%%%%%%%%%%%%%


\begin{thebibliography}{99}
%%%%%%%%%%%%%%%%%

\bibitem{SNIa}
A.~G.~Riess \textit{et al.}
%[Supernova Search Team Collaboration],
%``Observational evidence from supernovae
%for an accelerating universe and a cosmological constant,''
Astron.\ J.\  {\bf 116}, 1009 (1998) [astro-ph/9805201];
S.~Perlmutter \textit{et al.}
%[Supernova Cosmology Project Collaboration],
%``Measurements of Omega and Lambda from
%42 high redshift supernovae,''
Astrophys.\ J.\  {\bf 517}, 565 (1999) [astro-ph/9812133].

\bibitem{review}
E.~J.~Copeland, M.~Sami and S.~Tsujikawa,
%``Dynamics of dark energy,''
Int.\ J.\ Mod.\ Phys.\ D {\bf 15}, 1753 (2006)
%doi:10.1142/S021827180600942X
[hep-th/0603057];
A.~Silvestri and M.~Trodden,
%``Approaches to Understanding Cosmic Acceleration,''
Rept.\ Prog.\ Phys.\  {\bf 72}, 096901 (2009)
%doi:10.1088/0034-4885/72/9/096901
[arXiv:0904.0024 [astro-ph.CO]];
S.~Tsujikawa,
%``Dark energy: investigation and modeling,''
%doi:10.1007/978-90-481-8685-3_8
arXiv:1004.1493 [astro-ph.CO];
T.~Clifton, P.~G.~Ferreira, A.~Padilla and C.~Skordis,
%``Modified Gravity and Cosmology,''
Phys.\ Rept.\  {\bf 513}, 1 (2012)
%doi:10.1016/j.physrep.2012.01.001
[arXiv:1106.2476 [astro-ph.CO]];
A.~Joyce, B.~Jain, J.~Khoury and M.~Trodden,
%``Beyond the Cosmological Standard Model,''
Phys.\ Rept.\  {\bf 568}, 1 (2015)
%doi:10.1016/j.physrep.2014.12.002
[arXiv:1407.0059 [astro-ph.CO]];
P.~Bull {\it et al.},
%``Beyond $^{\Lambda CDM}$: Problems, solutions, and the road ahead,''
Phys.\ Dark Univ.\  {\bf 12}, 56 (2016)
%doi:10.1016/j.dark.2016.02.001
[arXiv:1512.05356 [astro-ph.CO]].

\bibitem{quin}
Y.~Fujii, Phys.\ Rev.\ D {\bf 26}, 2580 (1982);
L.~H.~Ford,
%``Cosmological Constant Damping
%By Unstable Scalar Fields,''
Phys.\ Rev.\ D {\bf 35}, 2339 (1987);
C.~Wetterich, Nucl. \ Phys \ B. {\bf 302}, 668 (1988);
B.~Ratra and P.~J.~E.~Peebles,
%``Cosmological Consequences of a Rolling Homogeneous Scalar Field,''
Phys.\ Rev.\ D {\bf 37}, 3406 (1988);
T.~Chiba, N.~Sugiyama and T.~Nakamura,
%``Cosmology with x matter,''
Mon.\ Not.\ Roy.\ Astron.\ Soc.\  {\bf 289}, L5 (1997)
doi:10.1093/mnras/289.2.L5
%[astro-ph/9704199];
P.~G.~Ferreira and M.~Joyce,
%``Structure formation with a selftuning scalar field,''
Phys.\ Rev.\ Lett.\  {\bf 79}, 4740 (1997)
%doi:10.1103/PhysRevLett.79.4740
[astro-ph/9707286];
R.~R.~Caldwell, R.~Dave and P.~J.~Steinhardt,
%``Cosmological imprint of an energy component
%with general equation of state,''
Phys.\ Rev.\ Lett.\  {\bf 80}, 1582 (1998)
%doi:10.1103/PhysRevLett.80.1582
[astro-ph/9708069].

\bibitem{Chiba}
T.~Chiba, A.~De Felice and S.~Tsujikawa,
%``Observational constraints on quintessence:
%thawing, tracker, and scaling models,''
Phys.\ Rev.\ D {\bf 87}, 083505 (2013)
%doi:10.1103/PhysRevD.87.083505
[arXiv:1210.3859 [astro-ph.CO]];
S.~Tsujikawa,
%``Quintessence: A Review,''
Class.\ Quant.\ Grav.\  {\bf 30}, 214003 (2013)
%doi:10.1088/0264-9381/30/21/214003
[arXiv:1304.1961 [gr-qc]].

\bibitem{stensor}
L.~Amendola,
%``Scaling solutions in general nonminimal coupling theories,''
Phys.\ Rev.\ D {\bf 60}, 043501 (1999)
%doi:10.1103/PhysRevD.60.043501
[astro-ph/9904120];
J.~P.~Uzan,
%``Cosmological scaling solutions of nonminimally coupled scalar fields,''
Phys.\ Rev.\ D {\bf 59}, 123510 (1999)
%doi:10.1103/PhysRevD.59.123510
[gr-qc/9903004];
T.~Chiba,
%``Quintessence, the gravitational constant, and gravity,''
Phys.\ Rev.\ D {\bf 60}, 083508 (1999)
%doi:10.1103/PhysRevD.60.083508
[gr-qc/9903094];
N.~Bartolo and M.~Pietroni,
%``Scalar tensor gravity and quintessence,''
Phys.\ Rev.\ D {\bf 61}, 023518 (2000);
%doi:10.1103/PhysRevD.61.023518
F.~Perrotta, C.~Baccigalupi and S.~Matarrese,
%``Extended quintessence,''
Phys.\ Rev.\ D {\bf 61}, 023507 (1999)
%doi:10.1103/PhysRevD.61.023507
[astro-ph/9906066].

\bibitem{Brans}
C.~Brans and R.~H.~Dicke,
%``Mach's principle and a relativistic theory of gravitation,''
Phys.\ Rev.\  {\bf 124}, 925 (1961).

\bibitem{stensorw}
J.~Martin, C.~Schimd and J.~P.~Uzan,
%``Testing for w < -1 in the solar system,''
Phys.\ Rev.\ Lett.\  {\bf 96}, 061303 (2006);
R.~Gannouji, D.~Polarski, A.~Ranquet and A.~A.~Starobinsky,
%``Scalar-tensor models of normal and phantom dark energy,''
JCAP {\bf 0609}, 016 (2006);
L.~Perivolaropoulos,
%``Reconstruction of extended quintessence
%potentials from the SnIa gold dataset,''
JCAP {\bf 0510}, 001 (2005).

\bibitem{fRw}
L.~Amendola and S.~Tsujikawa,
%``Phantom crossing, equation-of-state singularities,
%and local gravity constraints in f(R) models,''
Phys.\ Lett.\ B {\bf 660}, 125 (2008)
%doi:10.1016/j.physletb.2007.12.041
[arXiv:0705.0396 [astro-ph]];
W.~Hu and I.~Sawicki,
%``Models of f(R) Cosmic Acceleration that Evade Solar-System Tests,''
Phys.\ Rev.\ D {\bf 76}, 064004 (2007)
%doi:10.1103/PhysRevD.76.064004
[arXiv:0705.1158 [astro-ph]];
A.~A.~Starobinsky,
%``Disappearing cosmological constant in f(R) gravity,''
JETP Lett.\  {\bf 86}, 157 (2007)
%doi:10.1134/S0021364007150027
[arXiv:0706.2041 [astro-ph]];
S.~Tsujikawa,
%``Observational signatures of f(R) dark energy models
%that satisfy cosmological and local gravity constraints,''
Phys.\ Rev.\ D {\bf 77}, 023507 (2008)
%doi:10.1103/PhysRevD.77.023507
[arXiv:0709.1391 [astro-ph]];
H.~Motohashi, A.~A.~Starobinsky and J.~Yokoyama,
%``Phantom boundary crossing and anomalous growth
%index of fluctuations in viable f(R) models of cosmic acceleration,''
Prog.\ Theor.\ Phys.\  {\bf 123}, 887 (2010)
%doi:10.1143/PTP.123.887
[arXiv:1002.1141 [astro-ph.CO]].

\bibitem{obsersig}
S.~M.~Carroll, I.~Sawicki, A.~Silvestri and M.~Trodden,
%``Modified-Source Gravity and Cosmological Structure Formation,''
New J.\ Phys.\  \textbf{8}, 323 (2006);
R.~Bean, D.~Bernat, L.~Pogosian, A.~Silvestri and M.~Trodden,
%``Dynamics of Linear Perturbations in f(R) Gravity,''
Phys.\ Rev.\  D \textbf{75}, 064020 (2007);
S.~Tsujikawa, K.~Uddin, S.~Mizuno, R.~Tavakol and J.~Yokoyama,
%``Constraints on scalar-tensor models of dark energy from
%observational and local gravity tests,''
Phys.\ Rev.\ D {\bf 77}, 103009 (2008)
%doi:10.1103/PhysRevD.77.103009
[arXiv:0803.1106 [astro-ph]];
R.~Gannouji, B.~Moraes, D.~F.~Mota, D.~Polarski,
S.~Tsujikawa and H.~A.~Winther,
%``Chameleon dark energy models with characteristic signatures,''
Phys.\ Rev.\ D {\bf 82}, 124006 (2010)
%doi:10.1103/PhysRevD.82.124006
[arXiv:1010.3769 [astro-ph.CO]].

\bibitem{Deffayet09}
C.~Deffayet, G.~Esposito-Farese and A.~Vikman,
%``Covariant Galileon,''
Phys.\ Rev.\  D {\bf 79}, 084003 (2009)
[arXiv:0901.1314 [hep-th]];
C.~Deffayet, S.~Deser and G.~Esposito-Farese,
%``Generalized Galileons: All scalar models whose curved
%background extensions maintain second-order field equations
%and stress-tensors,''
Phys.\ Rev.\  D {\bf 80}, 064015 (2009)
[arXiv:0906.1967 [gr-qc]].

\bibitem{Ostro}
M.~V.~Ostrogradski, Mem. Acad. St. Petersbourg VI 4,
{\bf 385} (1850).

\bibitem{Horndeski}
G.~W.~Horndeski, Int.\ J.\ Theor.\ Phys.\ {\bf 10},
363-384 (1974).

\bibitem{Horn2}
C.~Deffayet, X.~Gao, D.~A.~Steer and G.~Zahariade,
%``From k-essence to generalised Galileons,''
Phys.\ Rev.\ D {\bf 84}, 064039 (2011)
%doi:10.1103/PhysRevD.84.064039
[arXiv:1103.3260 [hep-th]];
T.~Kobayashi, M.~Yamaguchi and J.~'i.~Yokoyama,
%``Generalized G-inflation: Inflation with the most
%general second-order field equations,''
Prog.\ Theor.\ Phys.\  {\bf 126}, 511 (2011)
[arXiv:1105.5723 [hep-th]];
C.~Charmousis, E.~J.~Copeland, A.~Padilla and P.~M.~Saffin,
%``General second order scalar-tensor theory,
%self tuning, and the Fab Four,''
Phys.\ Rev.\ Lett.\  {\bf 108}, 051101 (2012)
[arXiv:1106.2000 [hep-th]].

%\cite{deRham:2011by}
\bibitem{deRham:2011by} 
  C.~de Rham and L.~Heisenberg,
  %``Cosmology of the Galileon from Massive Gravity,''
  Phys.\ Rev.\ D {\bf 84}, 043503 (2011)
  doi:10.1103/PhysRevD.84.043503
  [arXiv:1106.3312 [hep-th]];
  L.~Heisenberg, R.~Kimura and K.~Yamamoto,
  %``Cosmology of the proxy theory to massive gravity,''
  Phys.\ Rev.\ D {\bf 89}, 103008 (2014)
  doi:10.1103/PhysRevD.89.103008
  [arXiv:1403.2049 [hep-th]].


\bibitem{coGa}
C.~Deffayet, G.~Esposito-Farese and A.~Vikman,
%``Covariant Galileon,''
Phys.\ Rev.\  D {\bf 79}, 084003 (2009)
[arXiv:0901.1314 [hep-th]];
C.~Deffayet, S.~Deser and G.~Esposito-Farese,
%``Generalized Galileons: All scalar models whose curved
%background extensions maintain second-order field equations
%and stress-tensors,''
Phys.\ Rev.\  D {\bf 80}, 064015 (2009)
[arXiv:0906.1967 [gr-qc]].

\bibitem{Gum}
C.~Deffayet, A.~E.~Gumrukcuoglu, S.~Mukohyama and Y.~Wang,
%``A no-go theorem for generalized vector Galileons
%on flat spacetime,''
JHEP {\bf 1404}, 082 (2014).
%doi:10.1007/JHEP04(2014)082
[arXiv:1312.6690 [hep-th]].

\bibitem{Deffayet:2016von}
C.~Deffayet, S.~Mukohyama and V.~Sivanesan,
%``On p-form theories with gauge invariant second order field equations,''
arXiv:1601.01287 [hep-th].

\bibitem{Deffayet:2010zh}
C.~Deffayet, S.~Deser and G.~Esposito-Farese,
%``Arbitrary $p$-form Galileons,''
Phys.\ Rev.\ D {\bf 82}, 061501 (2010).
%doi:10.1103/PhysRevD.82.061501
[arXiv:1007.5278 [gr-qc]].

\bibitem{Heisenberg}
L.~Heisenberg,
%``Generalization of the Proca Action,''
JCAP {\bf 1405}, 015 (2014)
%doi:10.1088/1475-7516/2014/05/015
[arXiv:1402.7026 [hep-th]].

\bibitem{Tasinato}
G.~Tasinato,
%``Cosmic Acceleration from Abelian Symmetry Breaking,''
JHEP {\bf 1404}, 067 (2014)
%doi:10.1007/JHEP04(2014)067
[arXiv:1402.6450 [hep-th]];
G.~Tasinato,
%``A small cosmological constant from Abelian symmetry breaking,''
Class.\ Quant.\ Grav.\  {\bf 31}, 225004 (2014)
%doi:10.1088/0264-9381/31/22/225004
[arXiv:1404.4883 [hep-th]].

\bibitem{Allys}
E.~Allys, P.~Peter and Y.~Rodriguez,
%``Generalized Proca action for an Abelian vector field,''
JCAP {\bf 1602}, 004 (2016)
%doi:10.1088/1475-7516/2016/02/004
[arXiv:1511.03101 [hep-th]].

\bibitem{Jimenez16}
J.~Beltran Jimenez and L.~Heisenberg,
%``Derivative self-interactions for a massive vector field,''
Phys.\ Lett.\ B {\bf 757}, 405 (2016)
%doi:10.1016/j.physletb.2016.04.017
[arXiv:1602.03410 [hep-th]].

\bibitem{Horndeskivec}
G.~W.~Horndeski,
%``Conservation of Charge and the Einstein-Maxwell
%Field Equations,''
J.\ Math.\ Phys.\  {\bf 17}, 1980 (1976).
%doi:10.1063/1.522837

\bibitem{DeFelice16}
A.~De Felice, L.~Heisenberg, R.~Kase, S.~Mukohyama, S.~Tsujikawa and Y.~l.~Zhang,
%``Cosmology in generalized Proca theories,''
arXiv:1603.05806 [gr-qc].

\bibitem{Barrow}
J.~D.~Barrow, M.~Thorsrud and K.~Yamamoto,
%``Cosmologies in Horndeski's second-order
%vector-tensor theory,''
JHEP {\bf 1302}, 146 (2013)
%doi:10.1007/JHEP02(2013)146
[arXiv:1211.5403 [gr-qc]].

\bibitem{Jimenez}
J.~B.~Jimenez, R.~Durrer, L.~Heisenberg and M.~Thorsrud,
%``Stability of Horndeski vector-tensor interactions,''
JCAP {\bf 1310}, 064 (2013)
%doi:10.1088/1475-7516/2013/10/064
[arXiv:1308.1867 [hep-th]].

\bibitem{TKK}
G.~Tasinato, K.~Koyama and N.~Khosravi,
%``The role of vector fields in modified gravity scenarios,''
JCAP {\bf 1311}, 037 (2013)
%doi:10.1088/1475-7516/2013/11/037
[arXiv:1307.0077 [hep-th]].

\bibitem{Hull}
M.~Hull, K.~Koyama and G.~Tasinato,
%``A Higgs Mechanism for Vector Galileons,''
JHEP {\bf 1503}, 154 (2015)
%doi:10.1007/JHEP03(2015)154
[arXiv:1408.6871 [hep-th]];
M.~Hull, K.~Koyama and G.~Tasinato,
%``Covariantized vector Galileons,''
Phys.\ Rev.\ D {\bf 93}, no. 6, 064012 (2016)
%doi:10.1103/PhysRevD.93.064012
[arXiv:1510.07029 [hep-th]].

\bibitem{Li}
W.~Li,
%``A unifying framework for ghost-free Lorentz-invariant
%Lagrangian field theories,''
arXiv:1508.03247 [gr-qc].

\bibitem{Jimenez:2014rna}
J.~B.~Jimenez and T.~S.~Koivisto,
%``Extended Gauss-Bonnet gravities in Weyl geometry,''
Class.\ Quant.\ Grav.\  {\bf 31} (2014) 135002
%doi:10.1088/0264-9381/31/13/135002
[arXiv:1402.1846 [gr-qc]];
%\bibitem{Jimenez:2016opp}
J.~Beltran Jimenez, L.~Heisenberg and T.~S.~Koivisto,
%``Cosmology for quadratic gravity in generalized Weyl geometry,''
JCAP {\bf 1604}, 046 (2016)
%doi:10.1088/1475-7516/2016/04/046
[arXiv:1602.07287 [hep-th]].
  
\bibitem{Cherenkov} 
G.~D.~Moore and A.~E.~Nelson,
%``Lower bound on the propagation speed of gravity from 
%gravitational Cherenkov radiation,''
JHEP {\bf 0109}, 023 (2001)
%doi:10.1088/1126-6708/2001/09/023
[hep-ph/0106220];
R.~Kimura and K.~Yamamoto,
%``Constraints on general second-order scalar-tensor 
%models from gravitational Cherenkov radiation,''
JCAP {\bf 1207}, 050 (2012).
%doi:10.1088/1475-7516/2012/07/050

\bibitem{GWde} 
B.~P.~Abbott {\it et al.} [LIGO Scientific and Virgo Collaborations],
%``Observation of Gravitational Waves from a Binary Black Hole Merger,''
Phys.\ Rev.\ Lett.\  {\bf 116}, 061102 (2016)
%doi:10.1103/PhysRevLett.116.061102
[arXiv:1602.03837 [gr-qc]];
D.~Blas, M.~M.~Ivanov, I.~Sawicki and S.~Sibiryakov,
%``On constraining the speed of gravitational waves following GW150914,''
Pisma Zh.\ Eksp.\ Teor.\ Fiz.\  {\bf 103}, no. 10, 708 (2016)
%doi:10.7868/S0370274X16100039
[arXiv:1602.04188 [gr-qc]].
    
\bibitem{scvector}
A.~De Felice, L.~Heisenberg, R.~Kase, S.~Tsujikawa, Y.~l.~Zhang and G.~B.~Zhao,
%``Screening fifth forces in generalized Proca theories,''
Phys.\ Rev.\ D {\bf 93}, 104016 (2016)
%doi:10.1103/PhysRevD.93.104016
[arXiv:1602.00371 [gr-qc]].

\bibitem{Fleury}
P.~Fleury, J.~P.~B.~Almeida, C.~Pitrou and J.~P.~Uzan,
%``On the stability and causality of scalar-vector theories,''
JCAP {\bf 1411}, 043 (2014).
%doi:10.1088/1475-7516/2014/11/043
[arXiv:1406.6254 [hep-th]].

\bibitem{Beu}
F.~Beutler {\it et al.}  [BOSS Collaboration],
%``The clustering of galaxies in the SDSS-III Baryon Oscillation
%Spectroscopic Survey: Testing gravity with redshift-space
%distortions using the power spectrum multipoles,''
Mon.\ Not.\ Roy.\ Astron.\ Soc.\  {\bf 443}, 1065 (2014)
[arXiv:1312.4611 [astro-ph.CO]].

\bibitem{Ledo}
L.~Samushia {\it et al.},
%``The Clustering of Galaxies in the SDSS-III Baryon
%Oscillation Spectroscopic Survey (BOSS):
%measuring growth rate and geometry with anisotropic clustering,''
Mon.\ Not.\ Roy.\ Astron.\ Soc.\  {\bf 439}, 3504 (2014)
[arXiv:1312.4899 [astro-ph.CO]].

\bibitem{Eriksen}
E.~Macaulay, I.~K.~Wehus and H.~K.~Eriksen,
%``Lower Growth Rate from Recent Redshift Space Distortion Measurements
%than Expected from Planck,''
Phys.\ Rev.\ Lett.\  {\bf 111}, 161301 (2013)
[arXiv:1303.6583 [astro-ph.CO]].

\bibitem{Vik}
A.~Vikhlinin {\it et al.},
%``Chandra Cluster Cosmology Project III: Cosmological Parameter Constraints,''
Astrophys.\ J.\  {\bf 692}, 1060 (2009)
[arXiv:0812.2720 [astro-ph]].

\bibitem{Planck}
P.~A.~R.~Ade {\it et al.} [Planck Collaboration],
%``Planck 2015 results. XIII. Cosmological parameters,''
arXiv:1502.01589 [astro-ph.CO].
%%CITATION = ARXIV:1502.01589;%%

\bibitem{WMAP}
G.~Hinshaw {\it et al.} [WMAP Collaboration],
%``Nine-Year Wilkinson Microwave Anisotropy Probe (WMAP)
%Observations: Cosmological Parameter Results,''
Astrophys.\ J.\ Suppl.\  {\bf 208}, 19 (2013)
%doi:10.1088/0067-0049/208/2/19
[arXiv:1212.5226 [astro-ph.CO]].

\bibitem{Tsuji15}
S.~Tsujikawa,
%``Possibility of realizing weak gravity in redshift
%space distortion measurements,''
Phys.\ Rev.\ D {\bf 92}, 044029 (2015)
%doi:10.1103/PhysRevD.92.044029
[arXiv:1505.02459 [astro-ph.CO]].

\bibitem{DKT}
A.~De Felice, T.~Kobayashi and S.~Tsujikawa,
%``Effective gravitational couplings for cosmological
%perturbations in the most general scalar-tensor theories
%with second-order field equations,''
Phys.\ Lett.\ B {\bf 706}, 123 (2011)
%doi:10.1016/j.physletb.2011.11.028
[arXiv:1108.4242 [gr-qc]].

\bibitem{Amendola}
L.~Amendola, M.~Kunz, M.~Motta, I.~D.~Saltas and I.~Sawicki,
%``Observables and unobservables in dark energy cosmologies,''
Phys.\ Rev.\ D {\bf 87}, no. 2, 023501 (2013)
[arXiv:1210.0439 [astro-ph.CO]].

\bibitem{Bellini}
E.~Bellini and I.~Sawicki,
%``Maximal freedom at minimum cost: linear large-scale structure
%in general modifications of gravity,''
JCAP {\bf 1407}, 050 (2014)
[arXiv:1404.3713 [astro-ph.CO]].

\bibitem{Perenon}
L.~Perenon, F.~Piazza, C.~Marinoni and L.~Hui,
%``Phenomenology of dark energy: general features of large-scale perturbations,''
JCAP {\bf 1511}, 029 (2015)
%doi:10.1088/1475-7516/2015/11/029
[arXiv:1506.03047 [astro-ph.CO]].

\bibitem{deRham:2014naa} 
C.~de Rham, L.~Heisenberg and R.~H.~Ribeiro,
%``On couplings to matter in massive (bi-)gravity,''
Class.\ Quant.\ Grav.\  {\bf 32}, 035022 (2015)
%doi:10.1088/0264-9381/32/3/035022
[arXiv:1408.1678 [hep-th]].

\bibitem{deRham:2014fha} 
C.~de Rham, L.~Heisenberg and R.~H.~Ribeiro,
%``Ghosts and matter couplings in massive gravity, bigravity and multigravity,''
Phys.\ Rev.\ D {\bf 90}, 124042 (2014)
%doi:10.1103/PhysRevD.90.124042
[arXiv:1409.3834 [hep-th]];
L.~Heisenberg,
%``Quantum corrections in massive bigravity and new effective 
%composite metrics,''
Class.\ Quant.\ Grav.\  {\bf 32}, no. 10, 105011 (2015)
%doi:10.1088/0264-9381/32/10/105011
[arXiv:1410.4239 [hep-th]];
L.~Heisenberg,
%``Non-minimal derivative couplings of the composite metric,''
JCAP {\bf 1511}, 005 (2015)
%doi:10.1088/1475-7516/2015/11/005
[arXiv:1506.00580 [hep-th]].

\bibitem{Sorkin}
B.~F.~Schutz and R.~Sorkin,
%``Variational aspects of relativistic field theories,
%with application to perfect fluids,''
Annals Phys.\  {\bf 107}, 1 (1977).

\bibitem{DGS}
A.~De Felice, J.~M.~Gerard and T.~Suyama,
%``Cosmological perturbations of a perfect fluid and noncommutative variables,''
Phys.\ Rev.\ D {\bf 81}, 063527 (2010).

\bibitem{Bardeen}
J.~M.~Bardeen,
%``Gauge Invariant Cosmological Perturbations,''
Phys.\ Rev.\ D {\bf 22}, 1882 (1980).

\bibitem{KS}
H.~Kodama and M.~Sasaki
%``Cosmological Perturbation Theory,''
Prog.\ Theor.\ Phys.\ Suppl. {\bf 78}, 1 (1984).

\bibitem{Mukhanov}
V.~F.~Mukhanov, H.~A.~Feldman and R.~H.~Brandenberger,
Phys.\ Rept.\  {\bf 215}, 203 (1992).

\bibitem{BTW}
B.~A.~Bassett, S.~Tsujikawa and D.~Wands,
%``Inflation dynamics and reheating,''
Rev.\ Mod.\ Phys.\  {\bf 78}, 537 (2006)
%doi:10.1103/RevModPhys.78.537
[astro-ph/0507632].

\bibitem{Kaiser}
N.~Kaiser,
%``Clustering in real space and in redshift space,''
Mon.\ Not.\ Roy.\ Astron.\ Soc.\  {\bf 227}, 1 (1987).

\bibitem{Tegmark}
M.~Tegmark {\it et al.}  [SDSS Collaboration],
%``The 3-D power spectrum of galaxies from the SDSS,''
Astrophys.\ J.\  {\bf 606}, 702 (2004)
[astro-ph/0310725].

\bibitem{AKS}
L.~Amendola, M.~Kunz and D.~Sapone,
%``Measuring the dark side (with weak lensing),''
JCAP {\bf 0804}, 013 (2008)
%doi:10.1088/1475-7516/2008/04/013
[arXiv:0704.2421 [astro-ph]].

\bibitem{Boi}
B.~Boisseau, G.~Esposito-Farese, D.~Polarski and A.~A.~Starobinsky,
%``Reconstruction of a scalar tensor theory of gravity
%in an accelerating universe,''
Phys.\ Rev.\ Lett.\  {\bf 85}, 2236 (2000)
%doi:10.1103/PhysRevLett.85.2236
[gr-qc/0001066];
S.~Tsujikawa,
%``Matter density perturbations and effective gravitational
%constant in modified gravity models of dark energy,''
Phys.\ Rev.\ D {\bf 76}, 023514 (2007)
%doi:10.1103/PhysRevD.76.023514
[arXiv:0705.1032 [astro-ph]];
S.~Nesseris,
%``Matter density perturbations in modified gravity models
%with arbitrary coupling between matter and geometry,''
Phys.\ Rev.\ D {\bf 79}, 044015 (2009)
%doi:10.1103/PhysRevD.79.044015
[arXiv:0811.4292 [astro-ph]];
A.~De Felice, S.~Mukohyama and S.~Tsujikawa,
%``Density perturbations in general modified
%gravitational theories,''
Phys.\ Rev.\ D {\bf 82}, 023524 (2010)
%doi:10.1103/PhysRevD.82.023524
[arXiv:1006.0281 [astro-ph.CO]];
A.~De Felice, R.~Kase and S.~Tsujikawa,
%``Matter perturbations in Galileon cosmology,''
Phys.\ Rev.\ D {\bf 83}, 043515 (2011)
%doi:10.1103/PhysRevD.83.043515
[arXiv:1011.6132 [astro-ph.CO]].

\bibitem{DT12}
A.~De Felice and S.~Tsujikawa,
%``Cosmological constraints on extended Galileon models,''
JCAP {\bf 1203}, 025 (2012)
%doi:10.1088/1475-7516/2012/03/025
[arXiv:1112.1774 [astro-ph.CO]].

\bibitem{2dF}
W.~J.~Percival {\it et al.}  [2dFGRS Collaboration],
%``The 2dF Galaxy Redshift Survey: Spherical harmonics analysis 
%of fluctuations in the final catalogue,''
Mon.\ Not.\ Roy.\ Astron.\ Soc.\  {\bf 353}, 1201 (2004)
[astro-ph/0406513].

\bibitem{6dF}
F.~Beutler {\it et al.},
%``The 6dF Galaxy Survey: z \approx 0 measurement of the growth rate and sigma_8,''
Mon.\ Not.\ Roy.\ Astron.\ Soc.\  {\bf 423}, 3430 (2012)
[arXiv:1204.4725 [astro-ph.CO]].

\bibitem{Wiggle}
C.~Blake {\it et al.},
%``The WiggleZ Dark Energy Survey: the growth rate of cosmic
%structure since redshift z=0.9,''
Mon.\ Not.\ Roy.\ Astron.\ Soc.\  {\bf 415}, 2876 (2011)
[arXiv:1104.2948 [astro-ph.CO]].

\bibitem{SDSS}
L.~Samushia, W.~J.~Percival and A.~Raccanelli,
%``Interpreting large-scale redshift-space distortion measurements,''
Mon.\ Not.\ Roy.\ Astron.\ Soc.\  {\bf 420}, 2102 (2012)
[arXiv:1102.1014 [astro-ph.CO]].

\bibitem{BOSS}
B.~A.~Reid {\it et al.},
%``The clustering of galaxies in the SDSS-III Baryon Oscillation Spectroscopic Survey:
%measurements of the growth of structure and expansion rate at z=0.57 from anisotropic clustering,''
Mon.\ Not.\ Roy.\ Astron.\ Soc.\  {\bf 426}, 2719 (2012)
[arXiv:1203.6641 [astro-ph.CO]].

\bibitem{Torre}
S.~de la Torre {\it et al.},
%``The VIMOS Public Extragalactic Redshift Survey (VIPERS). Galaxy clustering
%and redshift-space distortions at z=0.8 in the first data release,''
Astron.\ Astrophys.\  {\bf 557}, A54 (2013)
[arXiv:1303.2622 [astro-ph.CO]].

\bibitem{Okumura}
T.~Okumura {\it et al.},
%``The Subaru FMOS galaxy redshift survey (FastSound). IV. 
%New constraint on gravity theory from redshift space distortions at $z\sim 1.4$,''
Publ.\ Astron.\ Soc.\ Jap.\  {\bf 68}, no. 3, 38 (2016)
%doi:10.1093/pasj/psw029
[arXiv:1511.08083 [astro-ph.CO]].

\end{thebibliography}
\end{document}